\documentstyle[11pt,aaspp4]{article}
\def\distrbn{distribution }

\def\rec{reconstruction }
\def\gau{Gaussianization }
 
\def\gal{galaxy }
\def\z{redshift }

\def\den{density }
\def\be{\begin{equation}}
\def\ee{\end{equation}}
\def\hmpc{{h^{-1}\;{\rm Mpc}}}
\def\zb{Zel'dovich-Bernoulli }
\def\zc{Zel'dovich-continuity }
\def\zel{Zel'dovich }

\begin{document}

{\title{RECOVERING THE PRIMORDIAL DENSITY FLUCTUATIONS:
 A COMPARISON OF METHODS}
\author{
\bf Vijay K. Narayanan and
Rupert A.C. Croft 
}
\affil{Department of Astronomy, The Ohio State University, Columbus, OH 43210;}
\affil{ vijay,racc@astronomy.ohio-state.edu}
\affil{ Email: vijay,racc@astronomy.ohio-state.edu}

\bigskip
\bigskip
\bigskip
\centerline{\bf ABSTRACT}
\medskip

We present a comparative study of different methods for 
reversing the gravitational evolution of a cosmological density field 
to recover the primordial fluctuations.
We test six different approximate schemes in all: linear theory,
the Gaussianization technique of Weinberg (1992), two different quasi-linear
dynamical schemes (Nusser \& Dekel 1992, Gramann 1993), a hybrid 
dynamical-Gaussianization method (Narayanan \& Weinberg 1998)
 and the Path Interchange Zel'dovich Approximation (PIZA) of 
Croft \& Gazta\~{n}aga (1997).
The final evolved density field from an N-body simulation constitutes our 
test case. 
We use a variety of statistical measures to compare the initial density 
field recovered from it to the true initial density field, 
using each of the six different schemes.
These include point-by-point comparisons of the density fields
in real space, and the individual modes in Fourier space, as well as
global statistical properties such as the genus, the PDF of the density,
 and the distribution of peak heights and their shapes. 
We find linear theory to be substantially less accurate than the other 
schemes, all of which reverse at least some of the non-linear effects of 
gravitational evolution even on scales as small as $3 \hmpc$. 
The Gaussianization scheme, while being robust and easy to apply, 
is the least accurate after linear theory. 
The two quasilinear dynamical schemes, which are based on Eulerian 
formulations of the Zel'dovich Approximation, give similar
results to each other and are more accurate than Gaussianization, 
although they break down quite drastically when used outside their range 
 of validity, the quasilinear regime.
The complementary beneficial aspects of the dynamical and the Gaussianization
schemes are combined in the Hybrid  method, which uses a dynamical scheme to 
account for the bulk displacements of mass elements and corrects for any 
systematic errors using Gaussianization. 
We find this reconstruction scheme to be more accurate and robust than 
either the Gaussianization or dynamical method alone.
The final scheme, the PIZA, performs substantially better than  the 
others in all point by point comparisons. 
The PIZA does produce an oversmoothed initial density field, 
with a smaller number of peaks than expected, but recovers the PDF 
of the initial density with impressive accuracy on scales as small as 
$3 \hmpc$.

\slugcomment{Submitted to ApJ, \date}

\newpage
\section{INTRODUCTION}

In the standard model of structure formation, the observed \gal
\distrbn arises from the growth by gravitational instability of small 
amplitude 
primordial \den fluctuations.
We  need to understand the characteristics of these
seed density perturbations if we are to trace the history of structure 
formation in the universe.
In the simplest inflationary model for the origin of these 
perturbations, the primordial density fluctuations on the 
astrophysically relevant scales arise from the quantum 
noise in the inflaton field (\cite{guth82}; \cite{hawking82};
\cite{starobinsky82}; \cite{bardeen83}).
Another class of models identifies these seed fluctuations with the 
topological defects that remain as the relics of high energy phase 
transitions (\cite{kibble76}).
We need to confront all these models  with observational data to 
decide which model, if any, can correctly predict the structure we observe.
Such a comparison is non-trivial because, while the theories predict the
 properties of the {\it primordial} fluctuations, the observations measure 
the properties of the {\it present} day density field.
Thus, the observed properties reflect the primordial properties after they
have  been  distorted by the non-linear gravitational instability process.
The conventional study of the structure formation process 
has focused on a rather indirect method.
In this usual approach, we gravitationally evolve a model for the 
initial density field  forward in time using N-body methods.
We then  compare the properties of the resulting mass distribution with 
those of the observed \gal \distrbn assuming either that galaxies trace 
mass or by using a specific plausible biasing prescription to select 
galaxies.
We can then either accept or reject the model of the initial density field
depending on how accurately the properties of the simulated galaxy 
\distrbn match the observations (\cite{defw85}).
This method requires that we densely explore the full parameter space 
of all the possible models for the initial density field, and therefore,
its applicability is limited both by the accuracy of our comparisons and 
by the amount of computational time.
Under these circumstances, it would be extremely useful if we could
 reverse the effects of the gravitational evolution and  recover
 the primordial density fluctuations directly from the 
observed \gal distribution, as we could then directly analyze its properties.
In this paper, we compare the accuracy of the different methods that 
have been proposed to reverse the effects of gravitational clumping and 
recover the primordial density fluctuations.

The time evolution of the mass density fluctuations in an expanding 
background universe is described by a second order differential equation 
that has both growing and decaying mode solutions (\cite{peebles80}).
A direct numerical integration of this differential equation backwards 
in time will fail because the decaying mode will amplify any residual noise
that is present in the final density field.
Therefore, any method for recovering the initial density fluctuations must 
solve this problem using some approximations regarding the 
growth of fluctuations or some plausible assumptions about the nature
of the initial density fluctuations.
In this paper, we classify the various \rec schemes that have been proposed 
 into three major categories depending on how they overcome 
this problem and reverse the effects of gravitational instability.
 The schemes in the first category are ``gravitational time 
machines'' that  attempt to run gravity {\it backwards} in time.
They treat the mass density field as a self gravitating pressureless
fluid of particles
and solve the  fluid  mass and/or momentum conservation equations 
using a Lagrangian
approximation for the particle trajectories.
The simplest methods in this scheme assume that the comoving trajectories of
the particles  are straight lines during gravitational evolution 
(the Zel'dovich approximation, \cite{zel70}).
This  is a reasonable first approximation because, in linear perturbation
 theory, the direction of the gravitational acceleration stays constant 
in time.
The Zel'dovich-Bernoulli equation, derived by  Nusser \& Dekel (1992)
 from the Euler momentum conservation equation, and the Zel'dovich-continuity
 equation derived by Gramann (1993) from the mass continuity equation 
(see also Nusser et al.\ 1991) fall in this category.
These two dynamical schemes describe the time evolution of the velocity
 potential and the gravitational potential respectively, using {\it first} 
order differential equations that have only growing mode solutions.
These equations can then be integrated backwards in time quite easily to 
recover the corresponding initial potential fields.
The initial density field follows from these potential fields from the 
relevant linear theory relations between these quantities, which we will 
describe in more detail in \S2.
Kolatt et al. (1996) used a modified version of the Zel'dovich-Bernoulli
 scheme to construct mock redshift catalogs of our cosmic neighborhood,
which can then be used to study the different biases and selection 
effects that complicate the analysis of galaxy \z and peculiar velocity 
surveys.

The Gaussianization mapping method of Weinberg (1992, hereafter W92) belongs 
to the second category of \rec methods.
It is based on the approximation that the rank order of the initial mass
density field, smoothed over scales of a few Mpc, is preserved 
under non-linear gravitational evolution.
It further assumes that the initial density fluctuations form a Gaussian
random field.
The method employs a monotonic mapping of the smoothed
final  density field to a smoothed initial mass
density field that has a Gaussian one-point probability 
distribution function (PDF).
This method was used by Weinberg (1989) to show that a structure as massive
 as the Perseus-Pisces supercluster (\cite{pp86}) can form from the 
gravitational instability of small amplitude Gaussian initial density
 fluctuations.
Although in this paper we explicitly assume a Gaussian form for the PDF,
this category of \rec methods would also naturally include any other schemes
which assume a purely local monotonic mapping between the final and initial
densities. 
Using the predicted evolution of the PDF under gravitational instability
(with, for example, the Local Lagrangian Approximation of 
Protogeros \& Scherrer [1997] or the Spherical Collapse Model of 
Fosalba \& Gazta\~{n}aga [1997]),
we can also explore the reconstruction of  non-Gaussian initial conditions.

Narayanan \& Weinberg (1998, hereafter NW98) proposed a
hybrid \rec scheme that combines the features of the two 
categories described above.
This method also assumes that the initial density fluctuation field is a
Gaussian random field.
As shown by NW98, this \rec scheme can recover the initial density field 
more accurately than either the Lagrangian dynamical schemes
of the first category or the Eulerian Gaussian mapping scheme of the 
second  category, provided that the true initial density field is indeed
a Gaussian random field.

The third category of methods, pioneered by Peebles (1989, 1990), treats the
gravitational instability problem as a two-point boundary value problem 
and solves for
the trajectories of the mass particles by minimizing the action integral.
Shaya, Peebles \& Tully (1995) used this technique to reconstruct the 
trajectories and the initial positions of the galaxies within $3000$ 
kms$^{-1}$  assuming that they started out with vanishingly small 
initial peculiar velocities.
This method is computationally intensive, and so far it has not been 
applied  to the reconstruction of the initial density field
over a cosmologically interesting volume.
Croft \& Gazta\~{n}aga (1997, hereafter CG97) demonstrated that the 
Zel'dovich Approximation  is the least action solution when the particle 
trajectories are approximated by rectilinear paths. 
This simplifying assumption (which sacrifices some of the potential 
accuracy of the least action approach) was used by CG97 as the basis for the
 Path Interchange Zel'dovich Approximation (PIZA) reconstruction method.
In this method, the \rec problem reduces to finding the straight 
line trajectories of all particles by satisfying the condition  that the 
total mean square particle displacement between the initial and final 
positions be a minimum.  

In this paper, we test six different \rec schemes that fall under these 
three categories,  with a view  to check which of these methods can 
accurately and robustly recover the initial \den fluctuation field.
The various schemes that we test are:
(1) Linear theory,
(2) the Zel'dovich-Bernoulli scheme,
(3) the Zel'dovich-continuity scheme,
(4) Gaussianization,
(5) the Hybrid method, and 
(6) the PIZA scheme.
We first gravitationally evolve a known initial density fluctuation 
field using an N-body simulation code.
We assume that the initial fluctuations form a Gaussian random field,
as predicted by the simplest inflationary models.
This Gaussian assumption is the simplest one among a wide class of 
assumptions, and there is observational evidence from both microwave
 background anisotropies (e.g., \cite{kogut96}; \cite{heavens98}) and 
galaxy clustering (e.g., \cite{wgm87}; \cite{ndy95}; \cite{chiu98})
 that the primordial mass density fluctuations form a Gaussian random field.
We recover the initial density field from the final, non-linear,
gravitationally evolved field, using all six of the \rec methods listed 
above.
We then use a variety of statistical measures to compare the 
local and the global properties of the true initial \den field with those 
of  the initial \den fields recovered by the different \rec methods.
We examine both point-by-point comparisons between the true and the 
recovered initial density fields and the ability of the \rec  methods to 
accurately recover the Gaussian nature of the initial density field.
These comparisons will enable us to  understand the relative performance 
of these \rec  methods and will be useful for the reconstruction
of the primordial fluctuation field from the density field traced by
redshift surveys. 

All the \rec schemes in the first and the second category are designed
 to derive the initial mass density field from a continuous final density 
field.
Moreover, since the schemes in the first category are based on the 
perturbation theory expansions of the density and velocity fields,
they break down in the strongly non-linear regions (characterized by
$\vert \delta \vert \gg 1$).
Therefore, in all these \rec schemes, we reconstruct the smoothed initial 
density field from a final density field that is smoothed with a Gaussian
filter, so that the resulting smoothed final density field does not have 
any strong non-linearities.
On the other hand, the PIZA scheme recovers the initial density field
starting from the final locations of all the mass particles.
Thus, in the PIZA scheme  alone, we smooth the density field 
{\it after} recovering the initial density field.

All the \rec schemes require the final mass density distribution, while it
is the galaxy \distrbn that is the observable quantity.
In the case of biased galaxy formation, the observed galaxy number
density fluctuations are not equal to the underlying mass density
fluctuations.
Of all these six schemes, the \gau and the hybrid \rec schemes can be 
adapted to reconstruct from biased  galaxy density fields in a 
straightforward manner (see W92 and NW98), on the assumption that the 
biasing is local, while the other schemes cannot be adapted so easily.
However, in this paper, we ignore the possibility of biased galaxy 
formation and focus purely on gravitational dynamics.
In this sense, our work is similar in spirit to the work of 
Coles et al. (1993) and Sathyaprakash et al. (1995), who compared the 
validity of different approximations for the forward 
evolution of a density field under gravitational instability.
We note here that we will only test those methods that  reconstruct
the initial density field from the final {\it density} field.
We will not test those techniques that use the final peculiar 
velocity field ( e.g., \cite{nd92}) as the input.

The outline of the paper is as follows. 
We describe the various \rec schemes in detail in \S2, and in \S3
we describe our test case, the density field from the output of
an N-body simulation.
In \S4, we describe the statistical properties of the six 
reconstructed density fields and compare them with those of the 
true initial density field.
We then discuss the performance of each of the \rec schemes and compare 
their relative advantages and shortcomings in \S5.
We also describe the potential problems that might be encountered during 
an actual \rec of the initial mass density fluctuations from
present day galaxy \z catalogs.

\section{RECONSTRUCTION SCHEMES}

The \rec schemes in the first category solve for the time evolution 
of the density field using perturbation theory expansions of the
density and velocity fields.
The growth of density contrasts,
$\delta({\bf x}) \ \equiv \left[ \rho({\bf x}) - \bar \rho \right]/\bar \rho,$
in an expanding universe
can be analyzed using the equations of  ideal fluid flow as long 
as the trajectories of individual fluid elements do not cross
(i.e, before any shell crossing).
Denoting the comoving distance by ${\bf x}$, the peculiar velocity 
by ${\bf v} = d{\bf x}/dD$, and the perturbed gravitational potential 
by $\phi_{g}$, these three equations are, in the case of a 
pressureless gravitating fluid (\cite{peebles80}; \cite{gr93}),
the mass continuity equation,
\be
\frac{\partial \delta}{\partial D} + {\bf \nabla \cdot v} + \delta {\bf \nabla \cdot  v} + 
({\bf v \cdot \nabla})\delta = 0,
\label{eqn:masscon}
\ee
the momentum conservation equation,
\be
\frac{\partial \delta}{\partial D} + {\bf (v \cdot \nabla)v} + 
\frac{3\Omega}{2f^{2}(\Omega)}\frac {({\bf v}+{\bf \nabla} \phi_{g})}{D} = 0,
\label{eqn:momcon}
\ee
and the Poisson equation,
\be
\nabla ^{2}\phi_{g} = \frac{\delta}{D}.
\label{eqn:poi}
\ee
In these equations, $\Omega$ is the cosmological density parameter, $H$ is the 
Hubble constant,  $D(t)$ is the linear growth factor, and 
$f(\Omega) = \dot{D}/HD \approx \Omega^{0.6}$ (\cite{peebles80}).

Although these three equations can be solved using a wide range of
assumptions, there are three simple approximate solutions that are useful for
reconstructing the primordial density fluctuations.
The first method  uses the  linear perturbation theory approximation,
 while the remaining two methods use the \zel approximation together 
with the assumption that the velocity field remains irrotational during 
gravitational evolution.
We describe these three methods below.

\subsection{\it Linear theory}

In linear perturbation theory, which is the simplest approximate
solution to the equations listed above, 
we assume that the density contrast ($\delta)$ and the peculiar velocities 
($v$) are small.
We can then neglect all the terms involving $\delta {\bf v}$ and $v^{2}$.
The mass continuity equation can then be trivially integrated over the 
linear growth factor to give
\be
\delta = -D\left( {\bf \nabla \cdot v}\right).
\label{eqn:lin}
\ee
In this approximation, all the density fluctuations grow at the same rate,
and the gravitational potential $\phi_{g}$ remains
 constant throughout the gravitational evolution.
If we Fourier transform both sides of equation~(\ref{eqn:lin}), we see that all
 the Fourier modes of the density field evolve at the same rate, proportional 
to $D(t)$, and that the different Fourier modes evolve independently of each
 other.
Further, by combining equation~(\ref{eqn:lin}) with equation~(\ref{eqn:poi}), 
we see that $\phi_{g} = \phi_{v}$, where the velocity potential $\phi_{v}$ is
defined by ${\bf v} = -{\bf \nabla} \phi_{v}$.
Thus, equation~(\ref{eqn:lin}) gives a simple prescription for recovering the 
initial density field in the linear theory approximation.

\subsection{\it Zel'dovich-Bernoulli method}

The linear theory approximation, being purely local, 
does not specifically account for the displacements of the mass 
particles during gravitational evolution.
An elegant approximation that addresses this issue is the \zel 
approximation (\cite{zel70}), in which the mass particles are assumed 
to move in straight lines during gravitational evolution.
In this approximation, the Eulerian comoving position ${\bf x(t)}$ of a
 mass particle at any time $t$ is given in terms of its initial Lagrangian 
position ${\bf q}$ by
\be
{\bf x}(t) = {\bf q} + D(t) \Psi({\bf q}).
\label{eqn:zelapprox}
\ee
The essential feature of this \zel approximation is that the 
displacement of the mass particle from its initial location is assumed to
 be separable into a product of two functions, one of which depends only on 
time $\left[ D(t) \right]$ and the other  only on the initial location
 $\left[ \Psi({\bf q}) \right]$.
Nusser \& Dekel (1992) used the \zel approximation and the 
Euler momentum conservation equation together with  the assumption
that the velocity field remains irrotational during gravitational evolution 
to derive a first order differential equation for the 
evolution of the velocity potential $\phi_{v}$:
\be
\frac{\partial \phi_{v}}{\partial D} = \frac{1}{2} \vert \nabla \phi_{v}\vert
^{2}.
\label{eqn:zelber}
\ee
This equation, called the Zel'dovich-Bernoulli equation, can be easily
integrated backwards in time from the present epoch to the initial epoch 
(defined by $D(t_{i}) = 0$) to derive the initial velocity potential.
Since this equation evolves the velocity potential backwards in time, 
this \rec scheme is best suited to recovering the initial density field
from the present day peculiar velocity field.
However, after studying N-body simulations, Nusser et al. (1991) suggested 
the use of the following empirical relationship between the velocity field 
and the density field in the quasi-linear regime of gravitational instability:
\be
\nabla \cdot {\bf v} = -\left(\frac{\delta}{1+0.18\delta}\right).
\label{eqn:vdndbb}
\ee
Thus, given the final density field, we can form the velocity divergence 
field using equation~(\ref{eqn:vdndbb}) and then compute the final 
velocity potential $\phi_{v}$ from it using the relation
\be
\nabla^{2}\phi_{v} = -\nabla \cdot {\bf v}.
\label{eqn:vphiv}
\ee
Once we recover the initial velocity potential, we can use the fact that
 $\phi_{g} = \phi_{v}$ in the linear regime and derive the initial density
 field from $\phi_{g}$ using the Poisson equation.

\subsection{\it Zel'dovich-continuity method}

Gramann (1993) showed that the initial gravitational potential is more 
accurately recovered using the Zel'dovich-continuity 
equation of Nusser et al. (1991), which combines the \zel approximation
with the mass continuity equation.
In this case, the time evolution of the gravitational potential is described 
by the equation
\be
\frac{\partial \phi_{g}}{\partial D} = \frac{1}{2} \vert \nabla \phi_{g}\vert
^{2} + C_{g},
\label{eqn:zelcon}
\ee
where $C_{g}$ is the solution of the Poisson type equation
\be
\nabla ^{2}C_{g} = \sum_{i=1}^{i=3}\sum_{j=i+1}^{j=3}\left[
  {\partial^2 \phi_g \over \partial x_i^2} 
  {\partial^2 \phi_g \over \partial x_j^2} - 
  {\left(\partial^2 \phi_g \over \partial x_i \partial x_j \right)^2} 
  \right] .
\label{eqn:cgdef}
\ee
The initial gravitational potential can be determined by integrating 
equation~(\ref{eqn:zelcon}) backwards in time to the initial epoch
(defined by $D(t_{i}) = 0$).
The initial density fluctuation field can then be derived from this
initial gravitational potential using the Poisson equation.

Both the Zel'dovich-Bernoulli and the Zel'dovich-continuity schemes
naturally account for the dynamical displacements of the density features
during gravitational evolution, albeit in an approximate way.
However, equation~(\ref{eqn:zelber}) and equation~(\ref{eqn:zelcon}) are both
valid only as long as the density fluctuations are in the 
linear or  quasi-linear regimes (defined by $\vert \delta \vert \leq 1$).
They do not robustly recover the initial density in regions of very 
high density when the present day structures are highly non-linear
($\vert \delta \vert \gg 1$).
Therefore, they require that the final density field be smoothed quite heavily 
to remove any gross non-linearities, before the dynamical evolution equations
are integrated backwards in time.
We should also note that these two schemes, like linear theory, require
a field rather than a distribution of particles or galaxies as their input.
In order to generate a grid of values for this field, some sort of mass
assignment procedure must be carried out, which necessarily entails a degree
of smoothing.

\subsection{\it Gaussianization}

The Gaussianization \rec method of W92 belongs to the second category
of \rec schemes.
It is based on the assumption, motivated by studying N-body simulations,
 that non-linear gravitational evolution preserves the rank order of the 
mass density field.
This means that the high density regions in the initial field become the 
high density regions in the final conditions, low density regions in the 
initial field become the voids in the final density field, and so on in 
between.
The method employs a monotonic mapping of the smoothed
final  density field to a smoothed initial mass
density field that has a Gaussian one-point distribution function.
By construction, this procedure imposes a Gaussian PDF
for the initial mass density field.
The high overdensities in extreme non-linear regions are mapped
to the positive tail of the Gaussian distribution, while the 
voids are assigned density values in the negative tail
(see Figure 3 in W92 for a graphical illustration of the  mapping method).
The Gaussianization scheme can robustly recover the initial density 
field even in those places where the present day density field is 
quite non-linear because it involves a straightforward mapping 
procedure.
Therefore, this method can be used to reconstruct the primordial
fluctuations from even mildly smoothed fields.
However, this procedure relies on the strong theoretical assumption
that the initial density fluctuations have a Gaussian PDF.
Moreover, since it maps the smoothed final galaxy density field 
to a smoothed initial mass density field at the same Eulerian position, 
it does not explicitly account for any bulk displacements of galaxies
 during gravitational evolution.
These displacements are typically quite small 
(of the order of a few Mpc) and therefore not fatal to the 
Gaussianization \rec procedure itself, but they do reduce its accuracy.

\subsection{\it Hybrid method}

NW98 proposed a hybrid \rec scheme that enjoys most of the desirable 
features of both the dynamical methods of the first category and the 
Gaussianization method of the second category.
When applying this scheme, we first evolve the mass density 
field backwards in time
using a modified implementation of the Zel'dovich-continuity scheme 
that is described by equation~(\ref{eqn:zelcon}).
When we integrate the gravitational potential backwards in time,
we use a smoother potential for the source term in the right hand side 
of equation~(\ref{eqn:zelcon}).
We derive this smoother potential from an extra smoothed final density field
and integrate this smoother potential backwards simultaneously with the 
higher resolution density field.
NW98 tested different values of the smoothing length while
deriving this smoother potential and found that a Gaussian smoothing of 
$R_{s} = 4 h^{-1}$Mpc led to the best recovery of the initial density 
field, when the final density field is smoothed with a Gaussian filter 
of radius $R_{s} = 3 h^{-1}$Mpc.
We then Gaussianize this recovered {\it initial} density field, thereby 
improving the recovery in the high density regions.
NW98 demonstrated that this method recovers the initial density field more
accurately and robustly than either the Gaussianization or the 
dynamical schemes alone.
 
\subsection{\it PIZA}

The third category of \rec methods comprises the schemes 
that are based on the least action principle.
This approach was pioneered by Peebles (1989, 1990), who reconstructed the 
trajectories of individual mass particles by minimizing the action
 integral.
The action integral is minimized subject to the constraint that 
the initial peculiar velocities of the mass particles should vanish.
As the number of galaxies becomes very large, a straightforward
 application of this method to reconstruct the initial fluctuations 
from galaxy \z surveys becomes very difficult. 
In fact, as stated in \S1, this reconstruction method has so far only 
been used to trace the formation history of the 
Local Supercluster (\cite{shaya95}).
Giavalisco et al. (1993) generalized the \zel approximation using a 
series expansion and combined it with the least action principle to 
derive a parametrization for the orbits of  mass particles.
Susperregi \& Binney (1994) adapted this scheme to a mass density field
that is defined on an Eulerian grid and tested it on one and two
dimensional Gaussian random fields.

In this paper, we choose to test the Path Interchange Zel'dovich
Approximation scheme of CG97.
This scheme essentially consists of a means of applying 
Zel'dovich Approximation dynamics directly 
to an evolved particle distribution and
recovering the initial positions and velocities of particles.
The Zel'dovich Approximation being the least action solution
when particle paths are straight lines, the PIZA scheme is probably the
simplest and easiest to apply of the least action based schemes.
It also appears  to be one of the most promising of the methods based 
on its potential applicability to 
catalogs with a large number of galaxies.
The particle-based Zel'dovich Approximation has been shown 
(e.g., Coles, Melott \& Shandarin 1993) to be one of the most accurate and
robust dynamical approximations for forward evolution of a density field.
CG97 showed that the PIZA scheme has similar accuracy to the forward
Zel'dovich Approximation
 when it comes to predicting particle velocities and displacements.
We can therefore hope that it will compare favorably with the
other methods in tests of their ability to recover the primordial
density fluctuations. 

 In order to apply  the PIZA algorithm to
an evolved density distribution (details are given in CG97), we must have
the final positions of particles, as the scheme is  Lagrangian. 
We make use of these, final, boundary conditions, and the initial boundary
conditions that the Universe was homogeneous and that the particles started
with zero velocity. 
We therefore choose a uniform arrangement such as a grid for the initial
 positions of the particles. 
Our task is now to connect each one of these initial positions to the 
correct final position. 
This can be done by using the constraint that the action be a minimum, 
which in this case reduces (see CG97) to the minimization of the sum 
total of the squares of the particle  displacements  
(from the initial to final position). 
We minimize this sum by starting from a random arrangement of paths 
joining initial and final positions, and interchanging the end points 
of pairs of paths if the new configuration leads to a reduction in the 
action. 
We carry out this procedure on random pairs of paths until a minimum in 
the action is reached. We then have a solution for the displacements
at each initial grid point. 
The initial density field is then given by equation~(\ref{eqn:lin}).

The six \rec schemes described above are all derived using different 
approaches and/or  assumptions.
We now systematically test  these different methods by reconstructing 
the smoothed initial mass density field from the same final mass 
distribution.
This will enable us to directly compare the ability of the different 
\rec schemes to reconstruct the various features of the primordial density
fluctuation field.
Since we would like to recover the initial density field in 
as much detail as possible, we will concentrate on 
reconstructions of the initial density field smoothed with a small 
Gaussian filter of radius 
$R_{s} = 3h^{-1}$Mpc.
At this level of smoothing, the final mass density field 
still contains many regions that are quite non-linear.
The schemes in the first category are all designed to work only in
the quasi-linear regime ($\vert \delta \vert \leq 1$) and may
fail in the extremely non-linear regions.
However, we still test the performance of these schemes at this small
smoothing scale so as to understand the nature and magnitude of this potential
failure.
We will also test the different \rec methods on final density fields
smoothed with Gaussian filters of progressively larger radii.
The PIZA scheme requires the final positions of the mass particles as its
input. 
So, for this scheme alone, we smooth the density field {\it after} the \rec
procedure and before comparing it with the smoothed true initial density field.

\section{GENERATION OF THE TEST DENSITY FIELD}

We will test the different \rec schemes on the \den field derived from an 
N-body simulation, for which we know the true initial density field 
{\it a priori}.
We first generate a random density field on a periodic cubical box
of side $L_{\rm box} = 200h^{-1}$Mpc.
We choose random  phases for the Fourier components of the density field so 
that the resulting field is a Gaussian random field.
We use the matter power spectrum form suggested by 
Efstathiou, Bond \& White (1992),
\be
P(k) = \frac{Ak}{\left[1+ \left[ ak+\left(bk\right)^{3/2}+\left(ck\right)^{2} 
\right]^{\nu} \right]^{2/\nu}},
\label{eqn:pkdef}
\ee
where $a = (6.4/\Gamma)h^{-1}$Mpc, $b = (3.0/\Gamma)h^{-1}$Mpc,
 $c = (1.7/\Gamma)h^{-1}$Mpc, $\nu = 1.13$ and $A$  is the normalization of 
the power spectrum.
This two parameter family of power spectra is  characterized by the 
amplitude $A$ and by the shape parameter $\Gamma$, which is equal to $\Omega_{0}h$
in cold dark matter models with a small baryon density and scale invariant 
initial density fluctuations.
We use $\Gamma = 0.25$, a value that is consistent with 
the observed clustering properties of different galaxy catalogs (\cite{pd94}).
We normalize the power spectrum so that the rms fluctuation in density 
in spheres of radius $8h^{-1}$Mpc ($\sigma_{8}$) is unity, in accordance 
with the value measured from optical galaxy redshift surveys (\cite{dp83}).
This rms fluctuation amplitude $\sigma_{8}$ is related to the power spectrum 
$P(k)$ by
\be
 {\sigma}^{2}_{8} = \int_{0}^{\infty} 4\pi k^{2}P(k)\tilde W^{2}(kR)dk,
\label{eqn:s8def}
\ee
where $\tilde W(kR)$ is the Fourier transform of a top hat filter of 
radius $R = 8\hmpc$.

We evolve this density field forward in time using a particle-mesh (PM) 
code written by Changbom Park.
This code is described and tested in Park (1990).
We use $100^{3}$ particles and a $200^{3}$ force mesh in this PM simulation.
We start the gravitational evolution from a redshift of $z = 23$ and follow
it to $z = 0$ in 46 equal incremental steps 
of the expansion scale factor $a(t)$.
We  form the final continuous mass density 
field by cloud-in-cell (CIC) binning (Hockney \& Eastwood 1981) 
the gravitationally evolved discrete mass distribution onto a $100^{3}$ grid.
We use a Fast Fourier Transform (FFT) to smooth this final density field,
relying on the fact that the boundary conditions are periodic.

\section{COMPARISON OF RECONSTRUCTION SCHEMES}

We recover the initial density field from the final simulation
density field described above, using all six schemes described in \S2.
Except for the PIZA scheme, we recover the smoothed initial density fields
from the final density fields that are smoothed with Gaussian filters of 
radii $R_{s} = 3, 5, 8 $ and $10h^{-1}$Mpc.
However, since we would like to accurately recover even the small scale 
structures in the initial density field, we will primarily focus on the
 density field recovered with a $3h^{-1}$Mpc Gaussian smoothing.
As there are cases in which there are large differences in the performance
of the different schemes at different smoothing scales, we also show the
 contour plots and point to point comparisons of the density
 fields recovered with a $10h^{-1}$Mpc Gaussian smoothing.
Since the PIZA scheme requires the final positions of all the mass particles,
we recover the initial density field from the final discrete mass \distrbn
and then smooth this recovered initial density field with a
Gaussian filter of appropriate radius before comparing it with the 
true smoothed initial density field.
Hence, unlike the other reconstruction schemes, the performance of 
the PIZA scheme does not depend on the extent to which the 
Gaussian smoothing and the gravitational evolution of the density field
 are commutative.

\subsection{\it Visual appearance}

Figure 1 shows the isodensity contours in a slice through the 
smoothed initial density fields.
The density fields are recovered from a final density field that is
smoothed with a Gaussian filter of radius $R_{s} = 3h^{-1}$Mpc.
The slices correspond to the density field in the region
$(x1,y1) = (50,50) h^{-1}$Mpc to $(x2,y2) = (150,150) h^{-1}$Mpc
at a z-coordinate of $50h^{-1}$Mpc.
The contour levels range from $-2\sigma$ to $+2\sigma$ in steps of $0.4\sigma$,
where $\sigma$ is the rms fluctuation of the smoothed density field.
The true initial density field is shown in panel (a).
Linear theory (panel b) does not account for the non-linear growth of 
structures at all, so panel (b) appears the same as the 
final smoothed density field.
The peaks in the reconstructed density field are thus higher and 
the voids are more sparsely populated compared to those in the true
 initial density field.
The Zel'dovich-Bernoulli and the Zel'dovich-continuity schemes 
(panels c and d) both recover the smoothed initial density field quite
 well in the moderately dense regions, but they fail drastically near the 
high density regions and the recovered voids are not as deep as those in 
the true initial density field.
The Gaussianization method (panel e) recovers the initial density field 
quite robustly even near the high density peaks.
However, although the contour shapes for the Gaussianized density field are
similar to that of the true initial density field on large scales, the 
structures in the Gaussianized density field are slightly 
shifted from their true locations.
This failure to reproduce the correct locations of corresponding structures
 is not obvious from this contour plot, but it will show up as an increased
 scatter in the scatter plot of the density fields that we will consider below.
The density fields recovered by the hybrid method  and the PIZA scheme 
(panels f and g respectively) are both dynamically accurate and quite robust 
in the high density regions.
We will quantify the superior reconstruction by  the hybrid and the PIZA 
schemes using the cross-correlation coefficient below.
We also note here that the structures in the density field recovered by 
the PIZA scheme  appear rather globular and isotropic
compared to those in the true density field.

\begin{figure}
\centerline{
\epsfxsize=\hsize
\epsfbox[18 144 592 718]{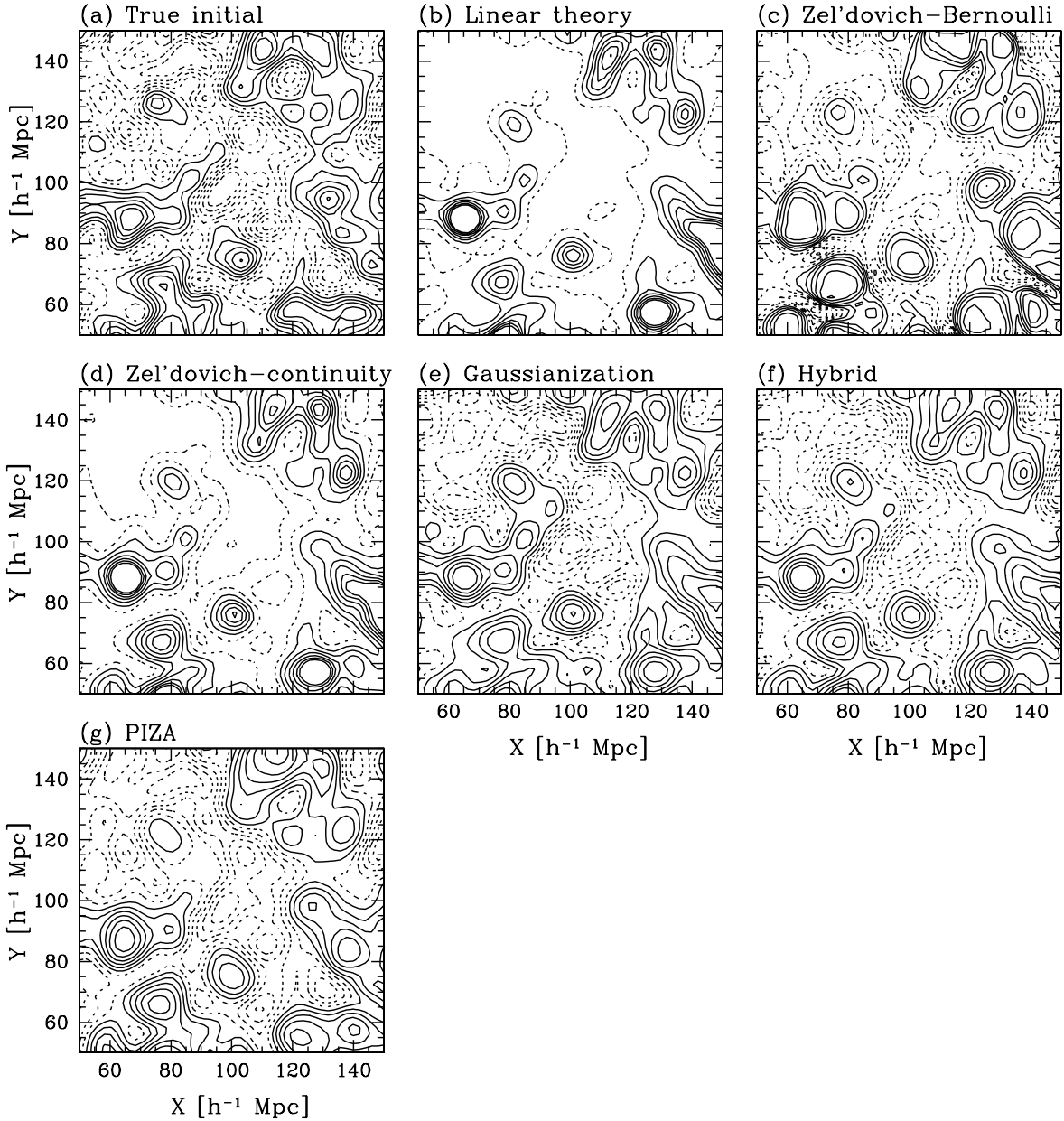}
}
\caption{Contours in a slice through the true and the reconstructed 
initial density fields.
The density fields are smoothed with a Gaussian filter of radius 
$R_{s} = 3h^{-1}$Mpc.
The contour levels range from $-2\sigma$ to $+2\sigma$ in steps of $0.4\sigma$.
Solid contours correspond to overdensities, while dashed contours
correspond to underdensities.
({\it a}) True initial conditions, a Gaussian random field  with a
 $\Gamma=0.25 $ power spectrum.
Remaining panels show the initial density field reconstructed from the final 
evolved density field by
({\it b}) Linear theory,
({\it c}) the Zel'dovich-Bernoulli scheme,
({\it d}) the Zel'dovich-continuity scheme,
({\it e}) Gaussianization,
({\it f}) the hybrid method and
({\it g}) PIZA.
}
\end{figure}

Figure 2 shows the true initial density field and the density fields
recovered from a final density field that is smoothed with a Gaussian 
filter of radius $R_{s} = 10h^{-1}$Mpc.
The format of this figure is identical to Figure 1 except that the contour
 levels range from $-2\sigma$ to $+2\sigma$ in steps of $0.2\sigma$.
At this large smoothing scale, the gravitationally evolved density field
is quite smooth, and as $\sigma$ is much lower (0.42 compared to 1.28
for $3 \hmpc$ smoothing), there are fewer highly non-linear structures.
Therefore, the dynamical schemes recover the smoothed initial field quite 
well even in the relatively high density regions.
 Linear theory is again the most inaccurate scheme, with the shallowness of
the voids being particularly noticeable. 
We will show below that the hybrid and the PIZA schemes still yield the 
most accurate recovery, although this superior performance is not
quite as evident from the contour plots.

\begin{figure}
\centerline{
\epsfxsize=\hsize
\epsfbox[18 144 592 718]{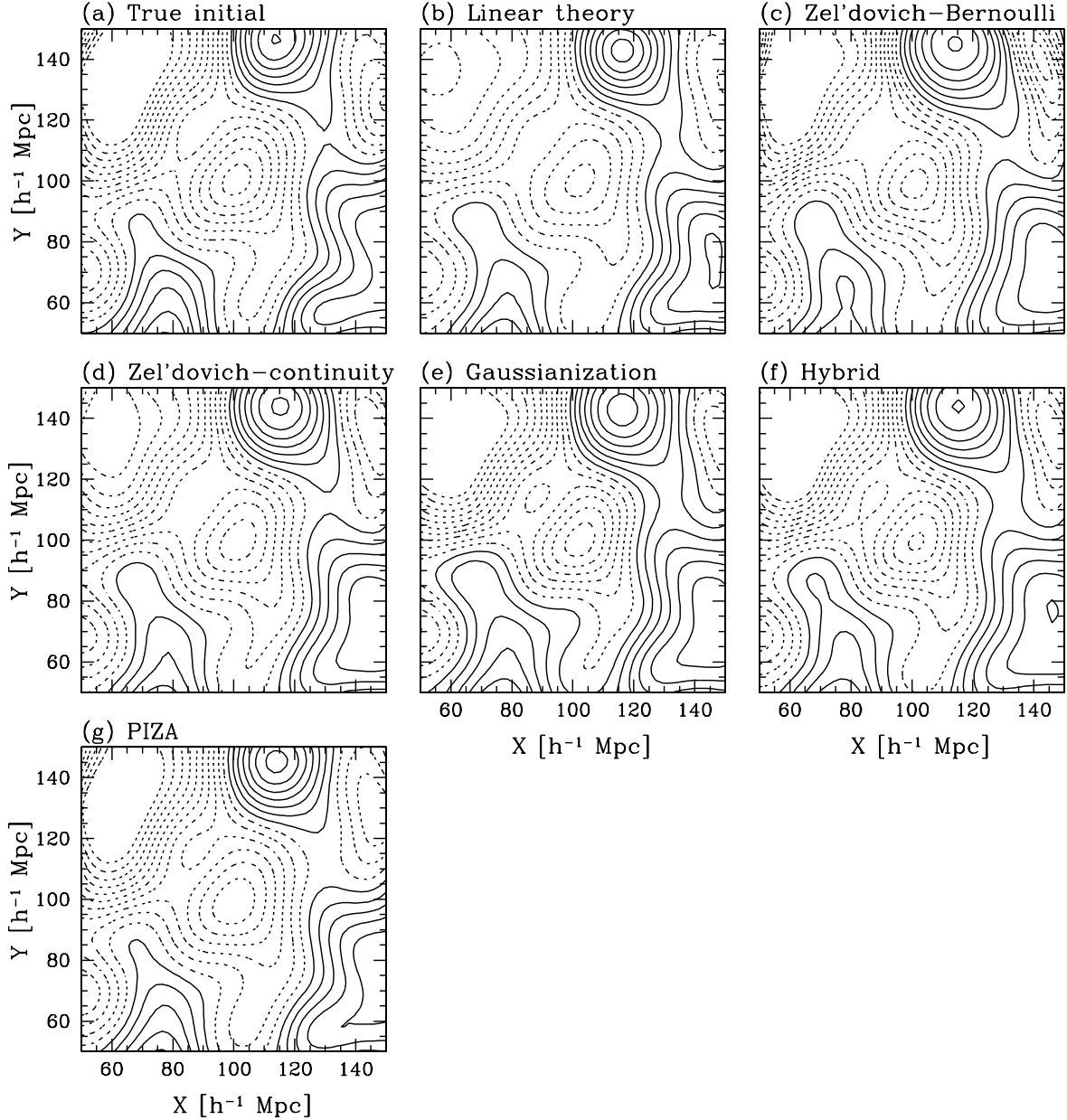}
}
\caption{Contours in a slice of the true and the recovered initial 
density fields in the same format as in Figure 1.
The density fields are smoothed with a Gaussian filter of radius 
$R_{s} = 10h^{-1}$Mpc.
The contour levels range from $-2\sigma$ to $+2\sigma$ in steps of $0.2\sigma$.
}
\end{figure}

\subsection{\it Point-by-point comparison}

Figure 3 shows plots of the scatter in a point-by-point 
comparison of the true and the reconstructed initial density fields
in 15625 cells.
We plot the density contrast at  cells in the reconstructed field  
($\delta_{r}$) against the true initial density contrast ($\delta_{i}$) at 
the same cells.
Each \distrbn  has been scaled by its RMS value so that the points in a 
perfect \rec lie on a straight line of unit slope.
We quantify the accuracy of the \rec  by the correlation coefficient $r$
between the reconstructed and  the true initial density fields,
\be
r = \frac{\left< \delta_{r}\delta_{i}\right>}{\left<\delta_{r}^{2}\right>^{\frac
{1}{2}}\left<\delta_{i}^{2}\right>^{\frac{1}{2}}}.
\label{eqn:rdef}
\ee
Linear theory yields the worst \rec of all the schemes.
The recovered initial \den field does not have any low \den regions at all
(with $\delta < -1$),
and there is a large scatter in the high \den regions.
The Zel'dovich-Bernoulli scheme recovers the  initial densities quite well,
with no systematic failures in the quasi-linear regions (characterized by
$\vert \delta \vert \leq 1$).
However, the relation exhibits noticeable curvature in the extremely 
overdense or underdense regions $(\vert \delta/\sigma \vert \geq 2)$.
The Zel'dovich-continuity scheme also clearly fails in the regions of large 
density contrasts.
It systematically overestimates the initial density in these regions and
produces a large scatter, resulting in a weaker correlation.
The Gaussianization scheme recovers the initial density field 
 in the highly non-linear regions without any systematic
 failures. 
However, its failure to account for the bulk displacements of the 
mass elements during gravitational evolution leads to a large scatter 
about the perfect \rec 
$\left[ ({\delta}/{\sigma})_{r} = ({\delta}/{\sigma})_{i} \right] $ and,
 consequently, a weak correlation.
The hybrid scheme corrects for these dynamical displacements using the 
Zel'dovich-continuity scheme, leading to a tighter correlation between the 
reconstructed and the initial density fields.
The PIZA scheme recovers the initial density field most accurately, 
without any systematic failures, and it exhibits the strongest correlation 
among all the \rec methods.

\begin{figure}
\centerline{
\epsfxsize=\hsize
\epsfbox[18 144 592 718]{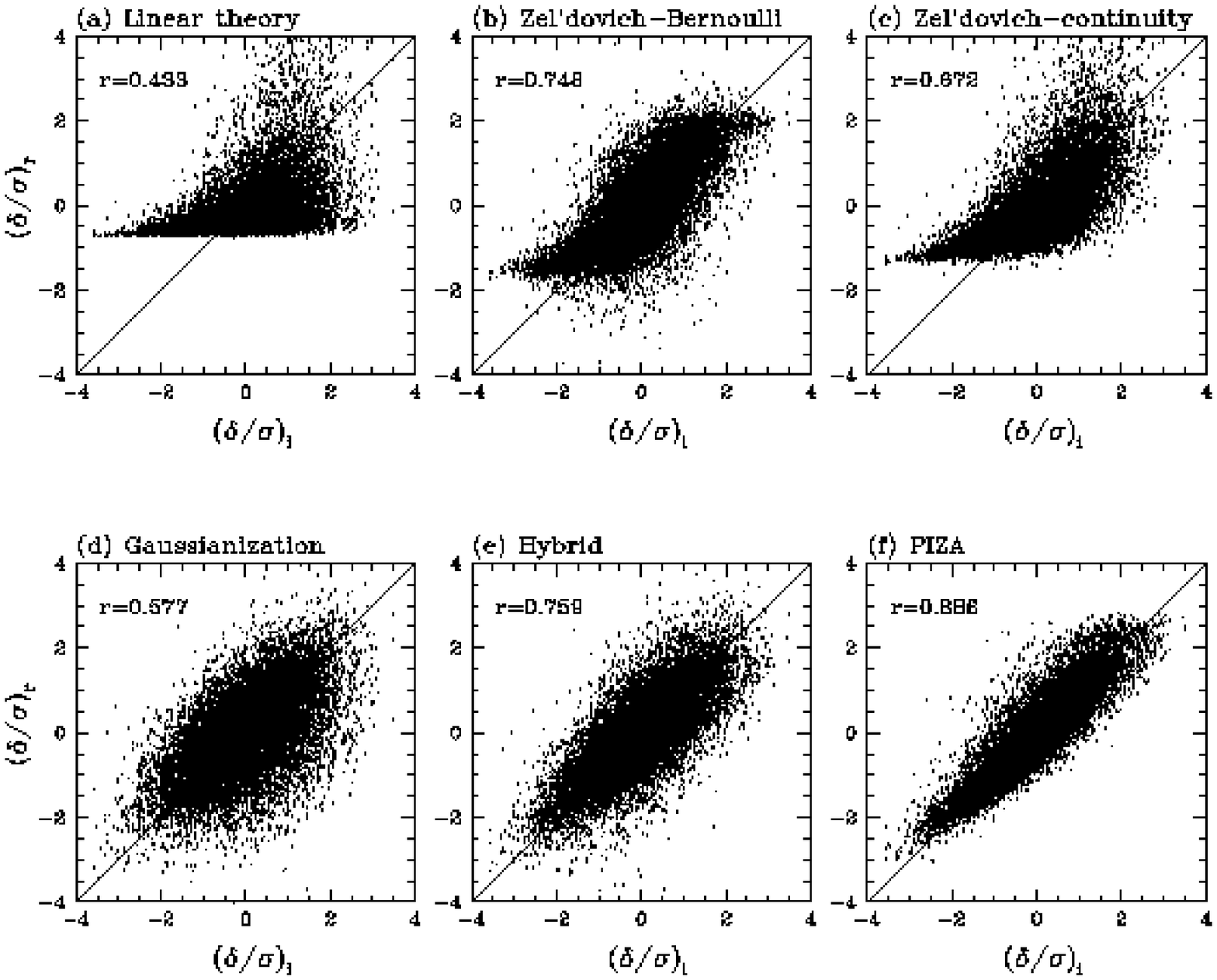}
}
\caption{ Cell by cell comparison of the recovered initial density contrast
$({\delta}/{\sigma})_{r}$ to the true initial density contrast 
$({\delta}/{\sigma})_{i}$.
The density fields are smoothed with a Gaussian filter of radius 
$R_{s} = 3\hmpc$.
The different panels correspond to the reconstruction using
({\it a}) Linear theory,
({\it b}) the Zel'dovich-Bernoulli scheme,
({\it c}) the Zel'dovich-continuity scheme,
({\it d}) Gaussianization,
({\it e}) the hybrid method and
({\it f}) PIZA.
The correlation coefficient between the two fields is indicated in each panel.
}
\end{figure}

Figure 4 shows the scatter plots for the reconstructions from a final 
density field that is smoothed with a Gaussian filter of radius
 $R_{s} = 10h^{-1}$Mpc.
When a large smoothing length such as this is used before
carrying out the reconstruction, the density contrasts are  mostly 
in the linear and quasi-linear regimes, so that the assumptions that go 
into formulating the Zel'dovich-Bernoulli and the Zel'dovich-continuity 
schemes are valid. 
These schemes do recover the initial density field at this level of 
smoothing fairly accurately, although some curvature of the type seen in the 
linear theory plot is also present in the  Zel'dovich-continuity results.
The \gau scheme does not show any curvature and has a good correlation, and
the hybrid \rec improves this correlation further by including a 
correction  for the displacements of density structures.
The PIZA scheme shows the tightest correlation at this smoothing scale
and has a visibly smaller scatter.

\begin{figure}
\centerline{
\epsfxsize=\hsize
\epsfbox[18 144 592 718]{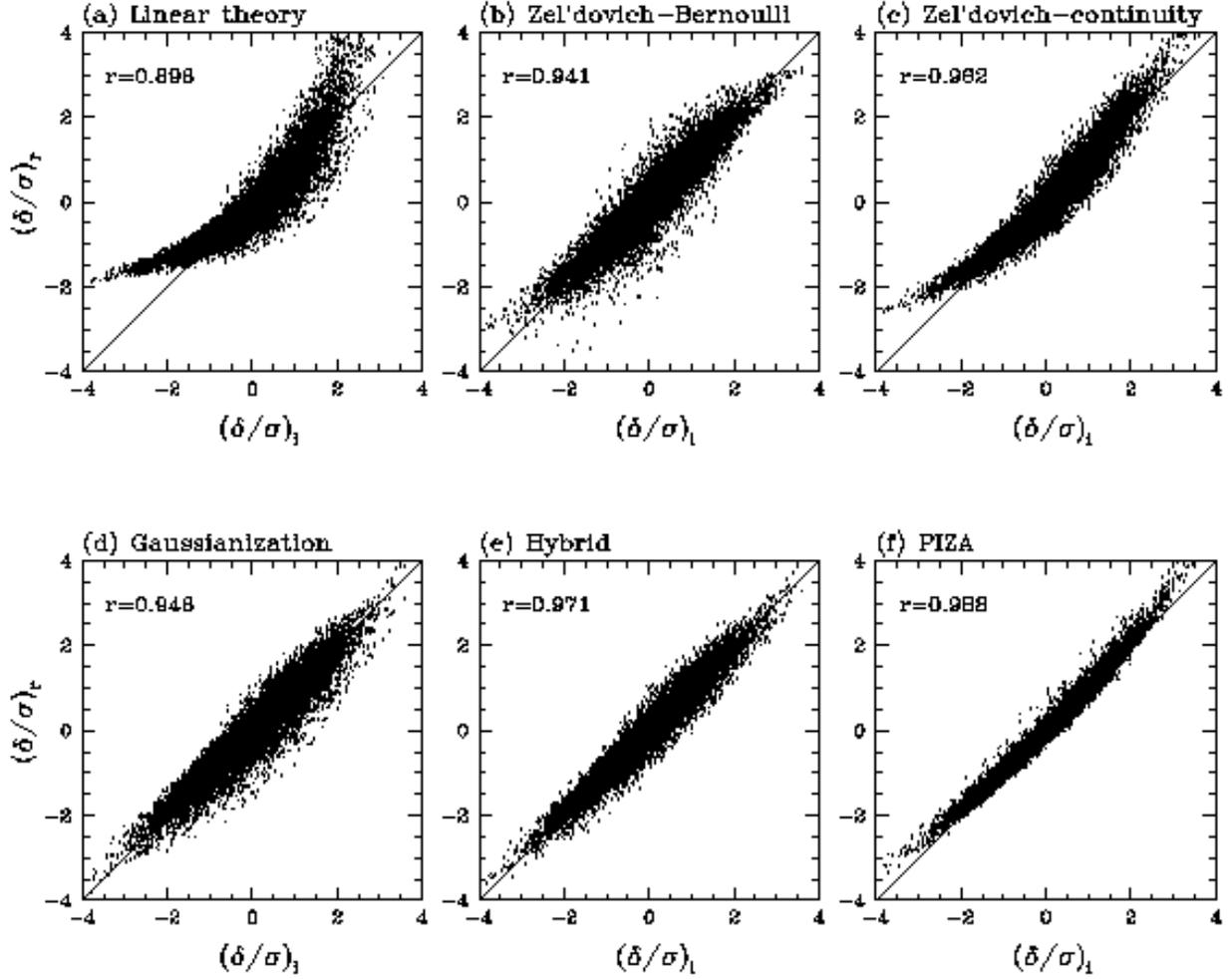}
}
\caption{ Cell by cell comparison of the recovered initial density contrast
$({\delta}/{\sigma})_{r}$ to the true initial density contrast 
$({\delta}/{\sigma})_{i}$ in the same format as in Figure 3.
All the density fields are smoothed with a Gaussian filter of radius 
$R_{s} = 10h^{-1}$Mpc  and scaled by their rms fluctuation $\sigma$. 
}
\end{figure}

\subsection{\it Reconstruction performance as a function of smoothing scale}

When we reconstruct the initial density fields with a small smoothing 
filter as in Figures 1 and 3, it is interesting to see how well the 
information  on larger scales is preserved.
 For example, we have seen that many of the schemes work rather
poorly when a point by point comparison is carried out between the
reconstructed and true initial fields without any additional smoothing.
To reach the results we show next (Figure 5), we have applied  additional
smoothing to the reconstructed and true initial density fields.
Figure 5 shows the correlation coefficient 
 as a function of the effective (total) Gaussian
 smoothing radius.
We first recover the initial density field from a final density field 
that is smoothed with a Gaussian filter of radius $R_{1} = 3h^{-1}$Mpc.
We then smooth this recovered initial field with another Gaussian filter of
radius $R_{2} = (R_{\rm eff}^{2} - R_{1}^{2})^{1/2}$, so that the recovered 
density field is smoothed, in effect, with a Gaussian filter of 
radius $R_{\rm eff}$.
This behavior will help us understand how accurately the information about 
the initial density field on {\it different} scales is 
recovered by the various 
\rec schemes, when they are used to reconstruct the initial density 
field smoothed on a {\it particular} scale.
This plot clearly shows that the PIZA scheme yields the tightest 
correlation between the true and the recovered initial density fields at 
all scales.
The hybrid scheme and the Zel'dovich-Bernoulli scheme recover the initial 
density field to almost the same accuracy at the different smoothing scales.
The Zel'dovich-continuity scheme is quite poor at recovering the small 
scale features and becomes progressively better in comparison with
the other schemes at larger scales.
The \gau scheme shows a weak correlation on all scales because of its
inherent Eulerian nature.
The linear theory scheme becomes relatively much worse as at the
smallest scales.

\begin{figure}
\centerline{
\epsfxsize=\hsize
\epsfbox[18 144 592 718]{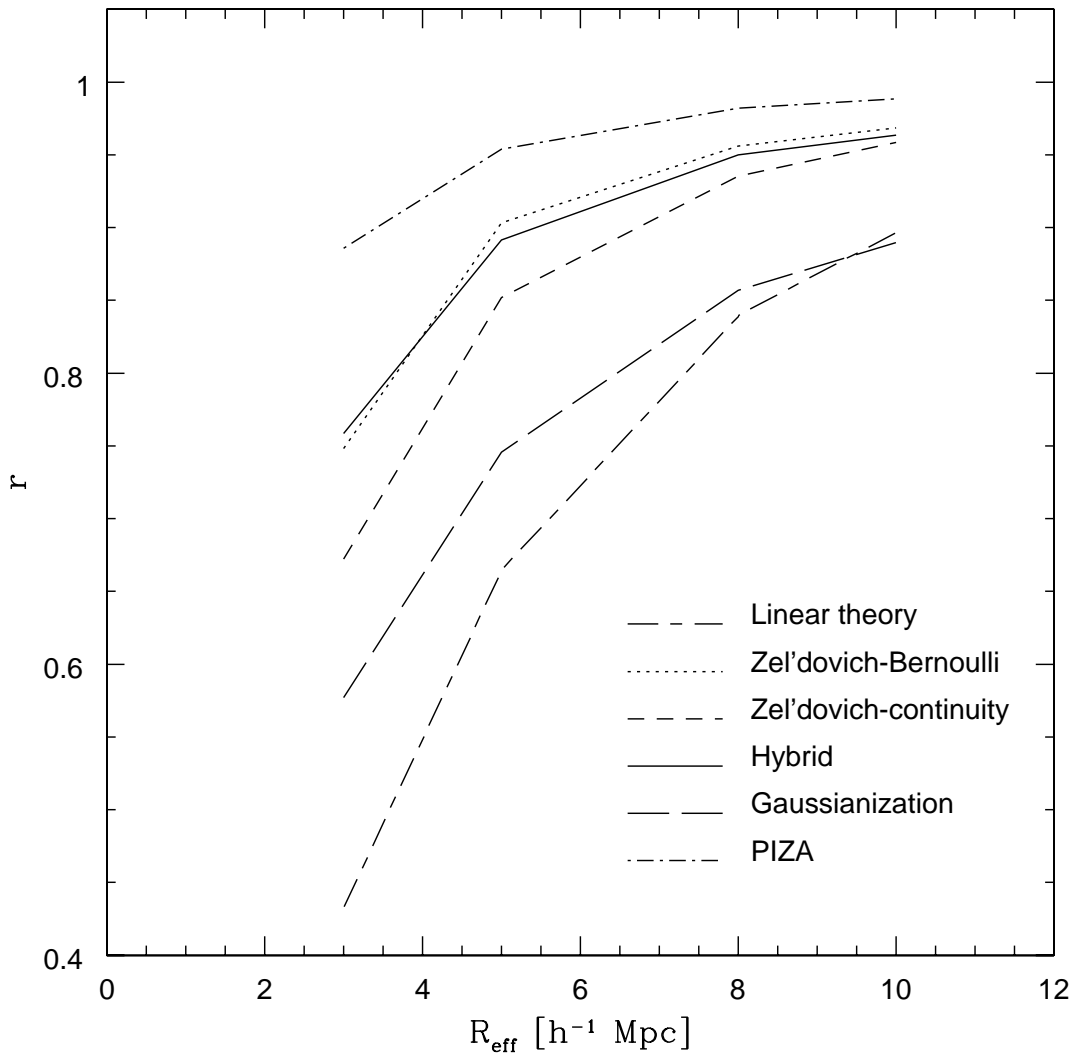}
}
\caption{Correlation between the true and the reconstructed initial density 
fields as a function of the effective Gaussian smoothing radius $R_{\rm eff}$,
for the various \rec schemes. 
The initial density fields are reconstructed from a final
density field that is smoothed with a $3\hmpc$ Gaussian filter.
The correlation coefficient is calculated after smoothing the recovered 
initial field further, giving a total effective smoothing $R_{\rm eff}$ 
(see \S4.3).
}
\end{figure}

In Figure 5, we showed how the information about the initial density field 
on different scales is recovered when it is reconstructed
from a final density field that is smoothed with a Gaussian filter of
radius $R_{s} = 3 \hmpc$. 
Alternatively, we can also reconstruct the initial density field 
on a particular smoothing scale from a final density field that is 
smoothed on the same scale.
This will enable us to quantify how accurately the different \rec schemes
can recover the initial density field on a {\it particular} smoothing scale.
For example, the difference between Figures 3 and 4 is the scale on which 
the final density fields were smoothed before reconstruction. 
We now extend the comparison shown in these figures to other scales and 
 show the correlation between the true and the recovered initial fields 
as a function of the scale of smoothing before reconstruction.
These results are presented in Figure 6.
 A comparison between Figures 5 and 6 indicates the extent to 
which the effects of gravitational evolution and Gaussian smoothing of 
the fields commute with each other.
The two procedures will give identical results for the linear theory and 
the PIZA \rec methods, because the linear theory \rec merely involves a
scaling of the amplitude of the final density field, while in the PIZA 
\rec scheme, all the smoothings are performed on the recovered
{\it initial} density field.
Figure 6  demonstrates the relative performance of the different \rec
schemes in the linear, quasi-linear and the non-linear regimes.
We see that even with heavy smoothing before 
reconstruction, none of the schemes can match the accuracy of the PIZA scheme.
The two grid-based Zel'dovich approximation schemes recover the 
initial density field with increasing accuracy as we go from the 
non-linear regime on small scales to the quasi-linear and linear regimes
on large scales.
We see that at large smoothing lengths the \gau scheme performs almost as 
well as the Zel'dovich schemes, presumably because the magnitude of the 
gravitational  displacements is now comparable to, or smaller than,
 the smoothing length of the final density field.

Comparing Figures 5 and 6, we see that the performance of the 
\zc scheme and the hybrid scheme are almost independent of the 
order of the smoothing and the \rec procedures.
On the other hand, the performance of the \zb and the \gau schemes
differs between the two cases.
The \gau \rec shows a higher correlation at any scale when the 
{\it final} density field is smoothed at the same scale.
However, the \zb scheme shows the opposite behavior, and yields a 
higher correlation when the {\it initial} density field recovered from
a mildly smoothed final density field is again smoothed using 
a larger Gaussian filter.
This is rather surprising, given that moderate density contrasts 
are supposedly a requirement of such a quasi-linear treatment. The
explanation is probably that the empirical correction (eq.~[\ref{eqn:vdndbb}]) 
to the final density field is most effective at relatively small 
smoothing lengths.

\begin{figure}
\centerline{
\epsfxsize=\hsize
\epsfbox[18 144 592 718]{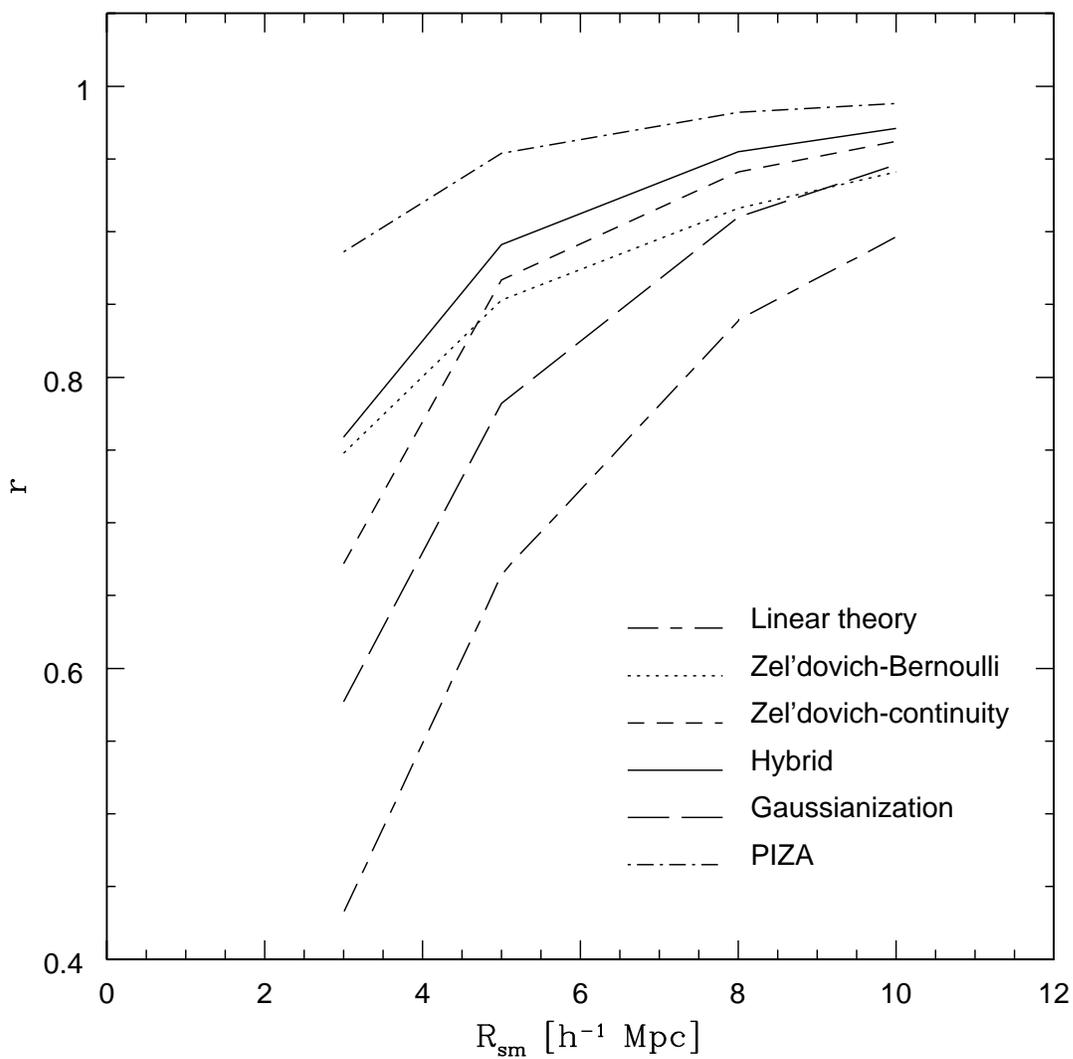}
}
\caption{Correlation between the true and the reconstructed initial density 
fields as a function of the radius $R_{\rm sm}$ of the Gaussian filter
used to smooth the final density field before reconstruction. 
}
\end{figure}

\subsection{\it Phase and amplitude correlations of the recovered
fields in Fourier space}

A natural measure of the relative accuracies of the reconstructions as a 
function of scale arises  in Fourier space, where we can quantify how well 
the different Fourier components of the true initial density field are 
recovered by the different  schemes.
Figure 7 shows the quantity
\be
D(k) = \frac{\sum \vert \tilde \delta_{r}({\bf k}) - \tilde \delta_{i}({\bf k}) \vert ^{2}}{\sum \left( \vert \tilde \delta_{r}({\bf k}) \vert ^{2} + \vert \tilde \delta_{i}({\bf k}) \vert ^{2}\right)},
\label{eqn:dkdef}
\ee
where the subscripts $i$ and $r$ refer to the true and the recovered initial 
density fields respectively.
The summation is over all the waves with wavenumbers in the interval 
$\left(k-k_{f},k\right]$, where  
$k_{f} = 2\pi/L_{\rm box} = 0.0314\  h\;$Mpc$^{-1}$ 
is the fundamental wavenumber of the simulation box of side 
$L_{\rm box} = 200h^{-1}$Mpc.
This statistic measures the difference in both the amplitudes and the 
phases of the Fourier components of the true and the recovered initial 
\den fields, and was first used by Little, Weinberg \& Park (1991)
to demonstrate the effects of power transfer from large scales
to small scales during non-linear gravitational evolution.
When the complex amplitudes of the Fourier components of the 
true and the recovered initial density fields are identical, $D(k) = 0$, 
while for two fields with uncorrelated phases, the average value of 
$D(k)=1$.
This quantity is independent of any smoothing of the density fields 
 after reconstruction and can test the ability of the different schemes to 
recover the Fourier components even below the smoothing scale.
The arrow marked $k_{\rm nl}$ in the Figure shows the wavenumber 
$k_{\rm nl} = 2\pi/2R_{\rm th} = 0.392\ h$Mpc$^{-1}$ that corresponds to
 the top-hat radius $R_{\rm th} = 8h^{-1}$Mpc, at which the rms amplitude
of density fluctuations is unity in linear theory.
This scale can be taken as a boundary between modes in the 
linear and non-linear regimes of gravitational evolution.
The large scale density modes with $k < k_{\rm nl}$ are still in the linear 
regime and evolve almost independently of each other, thereby retaining the 
phase information of the true initial density field.
On the other hand, the small scale modes with wavenumbers $k > k_{\rm nl}$
have all experienced phase shifts due to the strong coupling between
the evolution of the different modes in the non-linear stages of 
gravitational evolution (\cite{rg91}).
We find that the PIZA scheme recovers the Fourier modes of the true
 initial density field most accurately over a wide range of  scales.
The hybrid and the grid-based Zel'dovich dynamical schemes recover the 
initial Fourier components quite well up to the non-linear wavenumber 
$k_{\rm nl}$, but they fail at the larger  wavenumbers.
The \gau scheme is the only one which fails to recover the true phases
for the smallest wavenumbers, giving even worse results than linear theory. 
This behavior shows that the inaccuracies are not confined to small 
scales when the reconstruction smoothing scale is so small that 
the approximation of a monotonic Eulerian transformation between the initial
and the final density fields breaks down.

\begin{figure}
\centerline{
\epsfxsize=\hsize
\epsfbox[18 144 592 718]{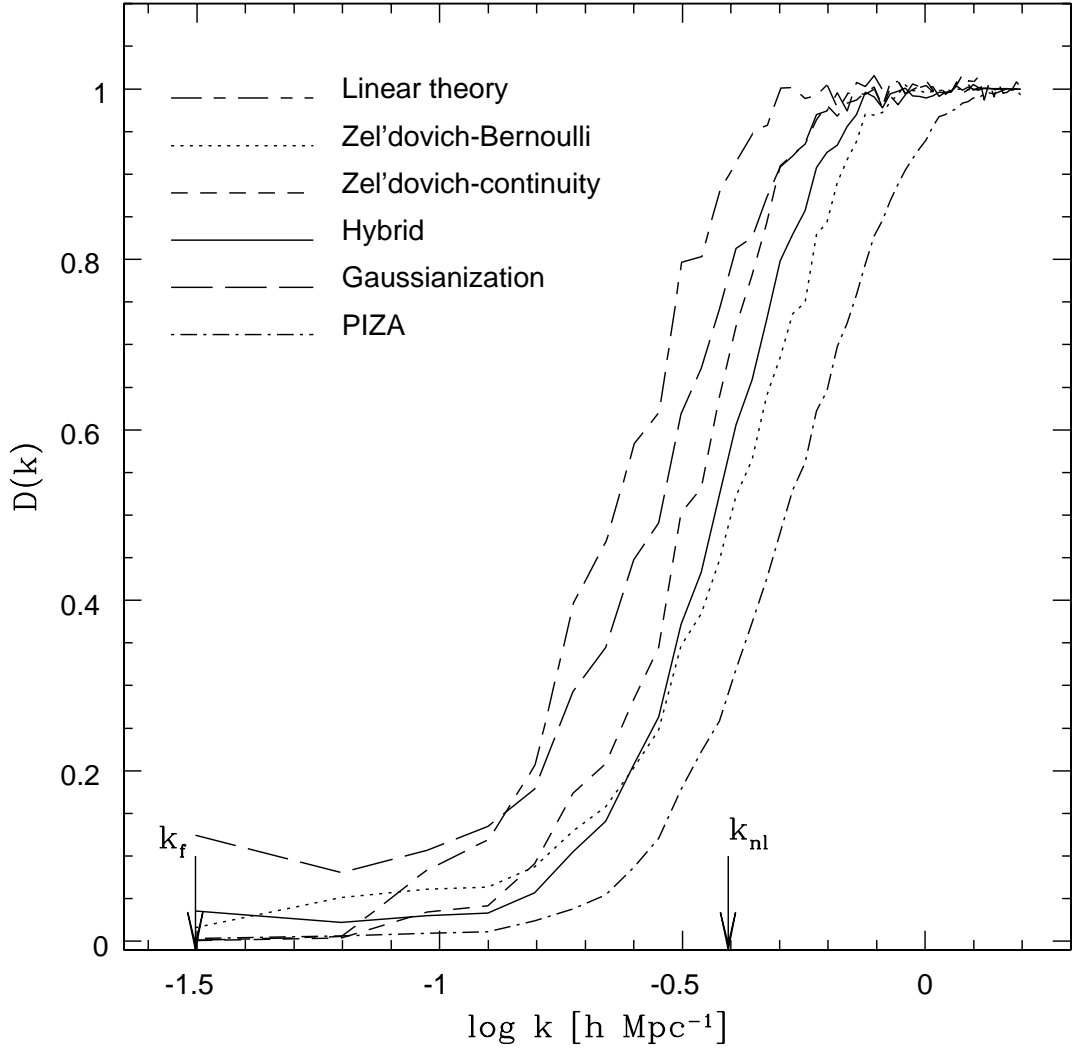}
}
\caption{Square of the difference between the complex
amplitudes of the Fourier components of the true and the recovered initial
 density fields, divided by the sum of their power spectra (see eq. [14]).
}
\end{figure}

\subsection{\it Probability distribution function}

In all the tests of the different \rec schemes we have considered so far, 
we have  focused on point-by-point comparisons between the true and 
the recovered initial density fields.
We now test how accurately the different schemes can recover the global 
statistical properties of the true initial density field, beginning with 
the one point PDF.

Figure 8 shows the PDF of the initial density field and the density field 
recovered by the various schemes from a final density field that is
 smoothed with a Gaussian filter of radius $R_{s} = 3h^{-1}$Mpc.
The true \den field has a Gaussian PDF by construction (solid points), while
the \gau and the hybrid \rec schemes explicitly impose a Gaussian PDF during 
the recovery procedure.
Non-linear evolution of the \den perturbations during the gravitational 
instability process induces a positive skewness in the PDF because,
 while the  overdensities can grow indefinitely with time, the underdensities 
cannot become more empty than $\delta = -1$.
Linear theory ignores this non-linear evolution and hence does not
restore the symmetry of the initial PDF.
The \zb and the \zc schemes are designed to reverse the effects of 
gravity in the linear and the quasi-linear regimes only.
The perturbation expansions on which they are based break down in
the very non-linear regions, with the result that they do not fully 
restore the symmetry between the positive and negative fluctuations of 
the true initial PDF.
The performance of the PIZA \rec scheme is impressive, 
because it {\it derives} the initial PDF from the final density field 
rather than imposing it by assumption.
The initial \den field recovered from the PIZA scheme does seem to be 
oversmooth, however, so that there are not enough highly overdense 
and highly underdense regions.
This will have some consequences for the properties  of peaks in the 
density field recovered by PIZA, as we will see below.

\begin{figure}
\centerline{
\epsfxsize=\hsize
\epsfbox[18 144 592 718]{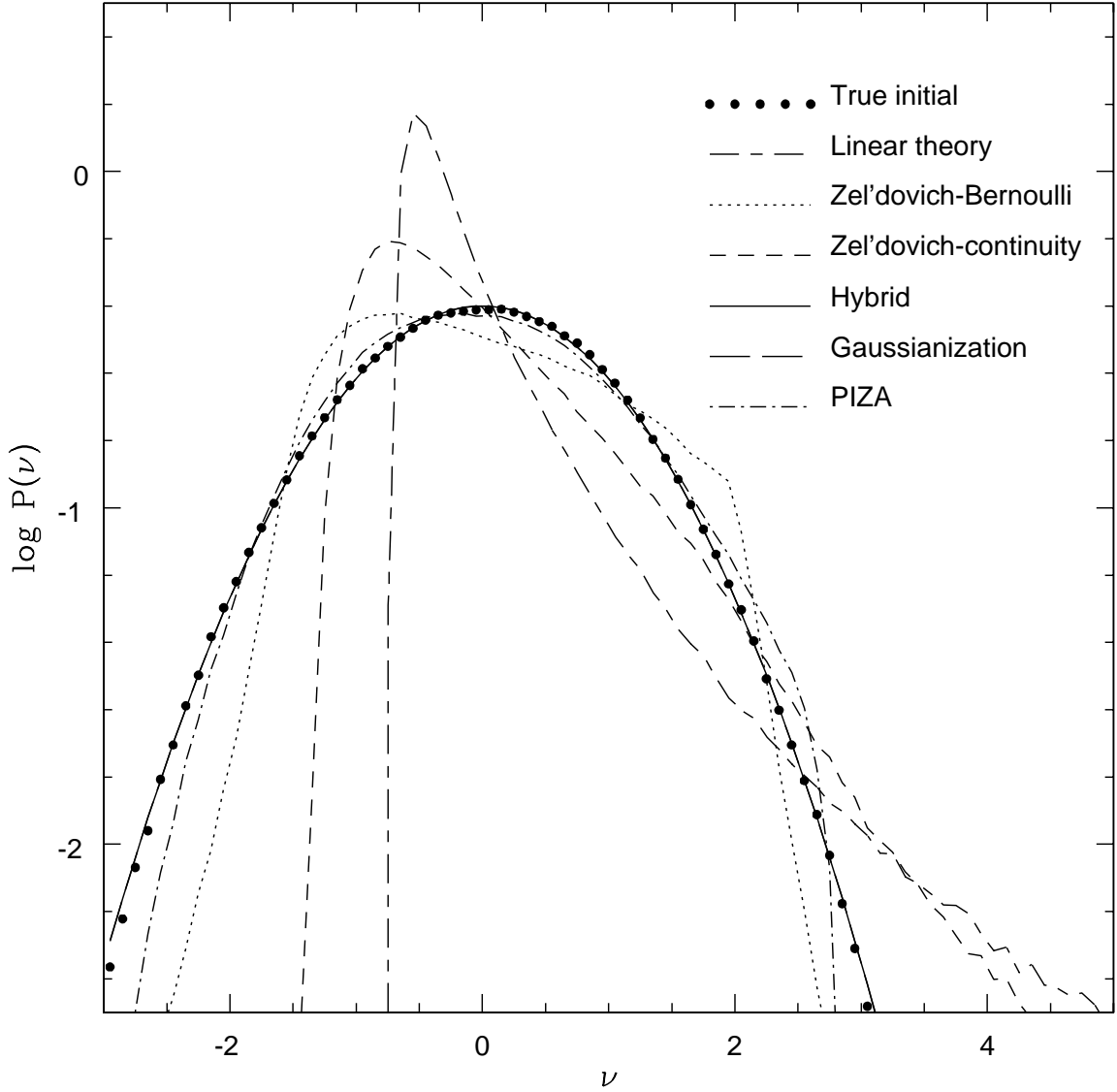}
}
\caption{ PDF of the true initial density field and the density field 
recovered by the six different \rec schemes for a Gaussian smoothing with 
$R_{s} = 3 \hmpc$.
The PDF of the Gaussianization \rec is covered by that of the hybrid 
reconstruction, as both the PDFs are exactly Gaussian by construction.
}
\end{figure}

\subsection{\it Topology}

Another global statistic is shown in Figure 9, the genus of the isodensity 
contour surfaces in the true and the recovered smoothed initial density 
fields as a function of the contour threshold density $\nu_{V}$.
The genus $G_{s}$ of a contour surface is defined as (\cite{wgm87}),
\be
G_{s} = ({\rm Number\ of\ holes}) - ({\rm Number\  of\  isolated\  regions}).
\label{eqn:gdef}
\ee
The contour threshold $\nu_{V}$ is defined implicitly in terms of the 
fraction ($f$) of the total volume that is enclosed by this isodensity 
contour as 
\be
f = (2\pi)^{-1/2} \int_{\nu_{V}}^{\infty} e^{-t^{2}/2}dt.
\label{eqn:fdef} 
\ee
For a Gaussian random field, $\nu_{V}$ is equal to the number of standard
deviations by which the threshold density differs from the mean density
(i.e, $\nu_{V} = \nu = \delta/\sigma$).
The true initial density field (filled circles) has the ``W'' shaped curve
 that is characteristic of a Gaussian random field 
(\cite{doroshkevich70}; \cite{adler81}; \cite{bbks}, hereafter BBKS; 
\cite{hgw86}).
Since Gaussianization preserves the rank order of the pixels,
it does not change the topology of the density field.
Thus, the genus curves of the density fields recovered using linear theory 
and the \gau \rec schemes are identical, as are the genus curves of the fields
reconstructed using the Zel'dovich-continuity scheme and the hybrid method.
Mildly non-linear gravitational evolution has  only a small 
effect on the shape of the genus curve, provided that the contour density
 threshold is defined in terms of the volume enclosed as in 
equation~(\ref{eqn:fdef}) (\cite{mwg88}; \cite{parkb91}).
 The shape of the genus curve of the density field reconstructed by  
linear theory (which is identical to the genus curve for the Gaussianized 
field) is therefore very similar to that of the true initial density 
field, although its amplitude is significantly smaller.
This amplitude drop arises due to strong phase correlations in the 
density field that develop during non-linear gravitational evolution,
and it has been observed in numerous studies of the non-linear evolution
of the genus curve  (\cite{mwg88}; \cite{parkb91}; \cite{springel98b}).
On the other hand, the genus curves for the density fields reconstructed by 
all the other methods show distinct shifts towards a ``meatball'' topology
(one dominated by isolated clusters).

\begin{figure}
\centerline{
\epsfxsize=\hsize
\epsfbox[18 144 592 718]{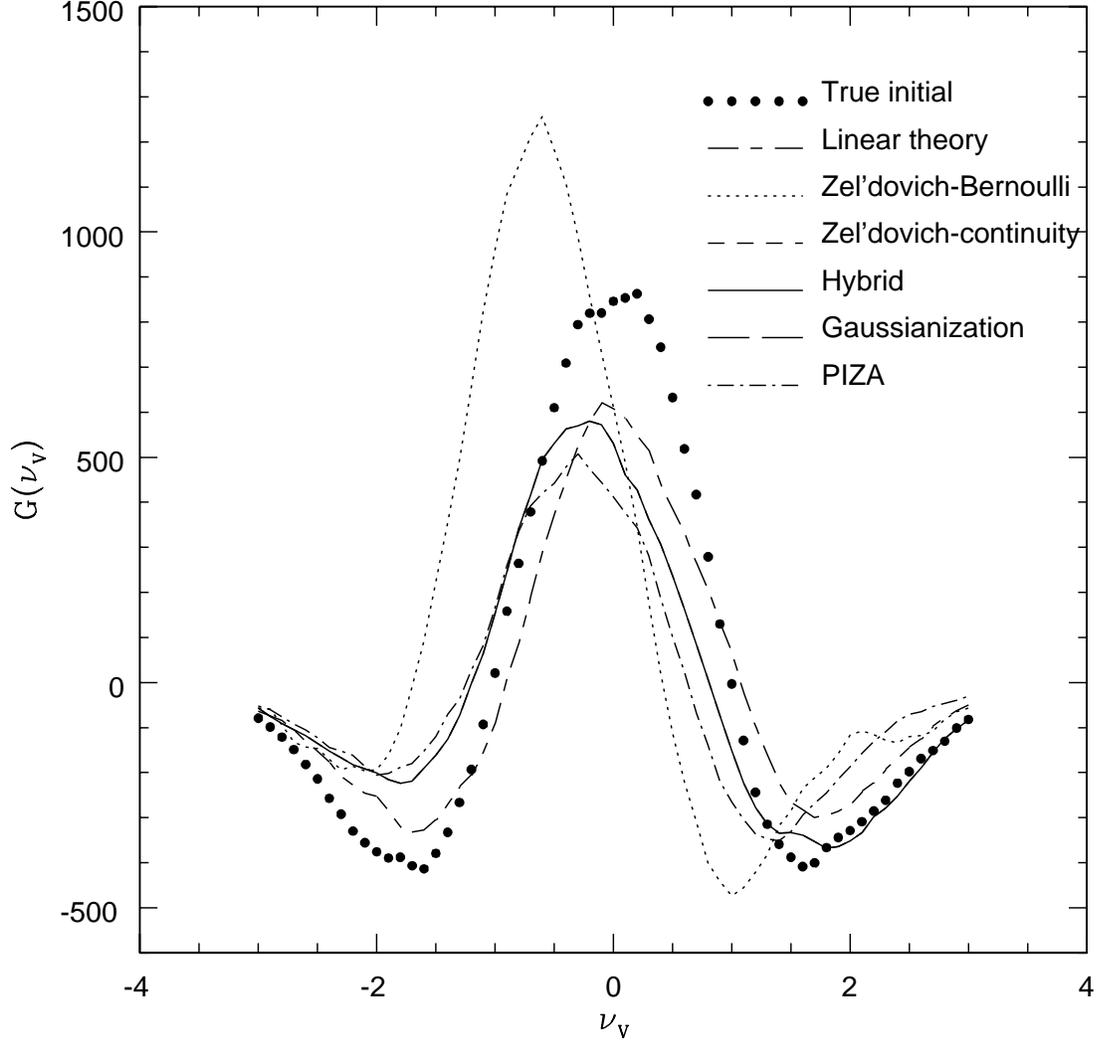}
}
\caption{ Genus curves of the true initial density field and the density field 
recovered by the different \rec methods for a Gaussian smoothing with 
$R_{s} = 3 \hmpc$.
The genus curves of the density field reconstructed using linear theory and 
the \gau scheme are identical, as are the curves for the 
density fields recovered
using the \zc and the hybrid schemes.
}
\end{figure}

\subsection{\it Peak heights and shapes}

Peaks in the initial density field are potential sites for the formation
of  galaxy clusters (\cite{kaiser84}; BBKS; \cite{colberg98}).
They are also the regions that undergo significant non-linear gravitational
collapse.
We now analyze how accurately the different \rec schemes can reproduce
the distribution of properties of the peaks in the true smoothed 
initial density field.
Figures 10 and 11 show the number of peaks in the true and the reconstructed 
initial density fields whose heights are greater than $\nu$ times the 
rms fluctuation $\sigma$ above the mean density 
(i.e, $\nu = \delta/\sigma$).
We identify the peaks as those pixels in the density field whose values are
higher than all their 26 neighboring pixels.
Figure 10 is a linear plot that clearly shows the differences between the
 different reconstructions for low values of $\nu$, while Figure 11 is a 
log plot that emphasizes the differences in the high $\nu$ region.
The filled triangles show this cumulative peak \distrbn for the true initial 
density field.
The solid circles show the number of peaks of different heights
expected in a Gaussian random field of the same volume. 
We compute this expected number using the equations for the peak number 
number density in \S4 of BBKS.
We see that, for $\nu < 2$, there are fewer peaks in the true 
field compared to the number predicted by BBKS.
We find this discrepancy to be due to the coarse resolution used in  the 
 CIC binning procedure.
The number of peaks for $\nu < 2$ becomes almost equal to the BBKS 
predicted number if we define the density field on a $200^{3}$ grid 
instead of on a $100^{3}$ grid.
However, all the reconstructed density fields will be affected in the same 
manner, so we can reliably compare the {\it relative} peak distributions 
of the different reconstructions with respect to that of the  true smoothed 
initial density field.

The \zb scheme recovers an excessive number of small and moderately high 
peaks, but it underestimates the number of very high peaks.
From visual inspection, we find that the extra peaks are located near 
the very high peaks in the true initial density field.
This, together with the fact that there is a deficiency in the number of
very high peaks, suggests that the extra peaks arise due to the failure 
of the \zb scheme near the highly overdense regions.
Thus, a single large peak in the final density field is broken 
down by the scheme into a large number of moderately high peaks.
This is also clear from the large number of ridge-like features 
surrounding the overdense regions in panel (c) of Figure 1.
Linear theory and the \zc scheme behave in the opposite 
manner, with an excessive number of very high peaks and a deficiency of 
moderate height peaks.
The \gau and hybrid schemes recover the peak distributions quite well,
as a result of  their robust performance in the very high density regions.
The PIZA scheme, on the other hand, severely underestimates the number of peaks
at all values of $\nu$, and there are no peaks whose heights are greater than 
 $ 3\sigma$ above the mean density in the recovered initial density field.

The PDFs of the initial density fields shown in Figure 8 suggest that at 
least part of the discrepancy in the peak number distribution can arise from 
differences in the PDFs themselves.
To check if the difference is due to some new failure of the \rec
procedures or merely the consequence of incorrect PDFs,
 we compare the peak number
distribution after imposing a common PDF on all the reconstructed initial 
density fields.
We Gaussianize the initial density fields recovered by all the \rec methods
other than the \gau and the hybrid schemes.
Figure 12 shows the peak number \distrbn after all the density fields
have been given the same Gaussian PDF.
The resulting density field for the linear theory \rec will be
 identical to the  density field reconstructed by Gaussianization, 
while the results for the \zc scheme will be identical to those for 
the hybrid reconstructed density field.
Therefore, we do not show the peak number \distrbn for these two 
reconstructions in this Figure.
The \zb \rec now agrees  well with the true initial peak number 
distribution, suggesting that the discrepancy seen in Figure 11 is largely 
due to an erroneous PDF (as shown in Figure 8).
The PIZA scheme now matches better with the true peak distribution at
the high peak height end, although there is still a large discrepancy
for peaks with  $\nu <3$, by as much as a factor of 3. 
This means that the rounding off of the PDF is probably not the dominant 
problem affecting the number of peaks in the PIZA scheme.
The slightly oversmoothed nature of the recovered density field seen in
 Figure 1g seems to make a number of smaller peaks disappear altogether.

\begin{figure}
\centerline{
\epsfxsize=\hsize
\epsfbox[18 144 592 718]{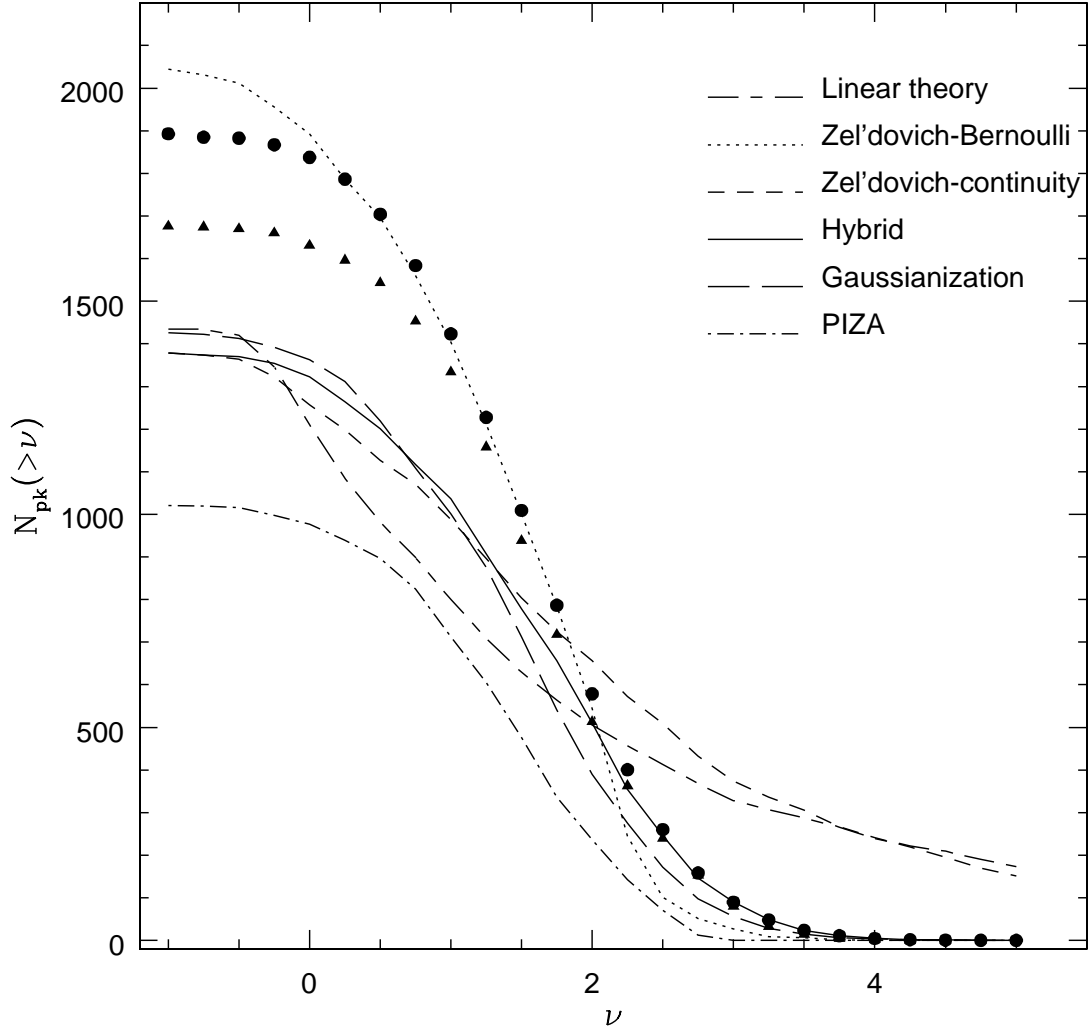}
}
\caption{ 
Number of peaks in the true and the reconstructed initial density fields 
whose heights are greater than $\nu\sigma$ above the mean density.
The filled circles show the number of peaks predicted by the BBKS formalism,
while the filled triangles show the number of peaks present in the true 
initial density field.
}
\end{figure}

\begin{figure}
\centerline{
\epsfxsize=\hsize
\epsfbox[18 144 592 718]{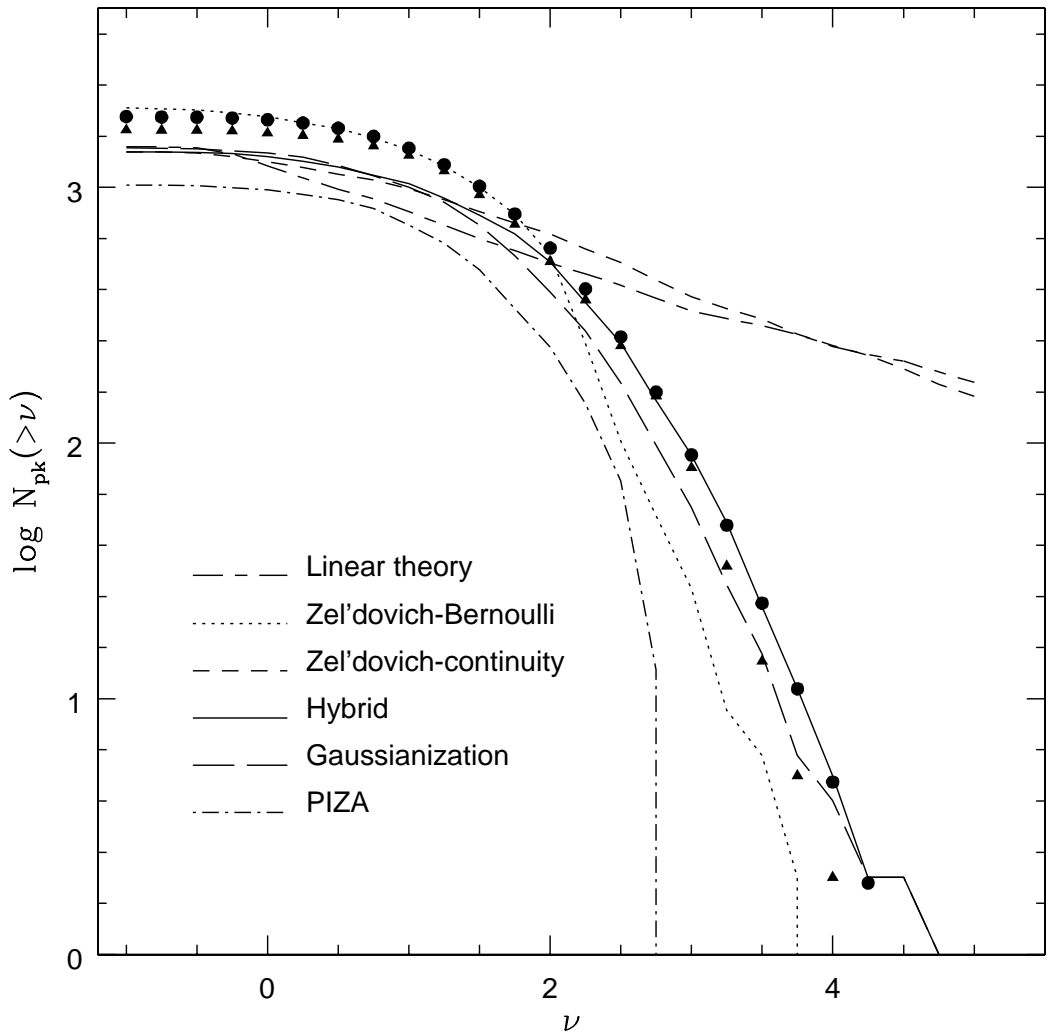}
}
\caption{ 
Same as Fig. 10, but with a logarithmic scale to emphasize the behavior of
high $\nu$ peaks.
}
\end{figure}

\begin{figure}
\centerline{
\epsfxsize=\hsize
\epsfbox[18 144 592 718]{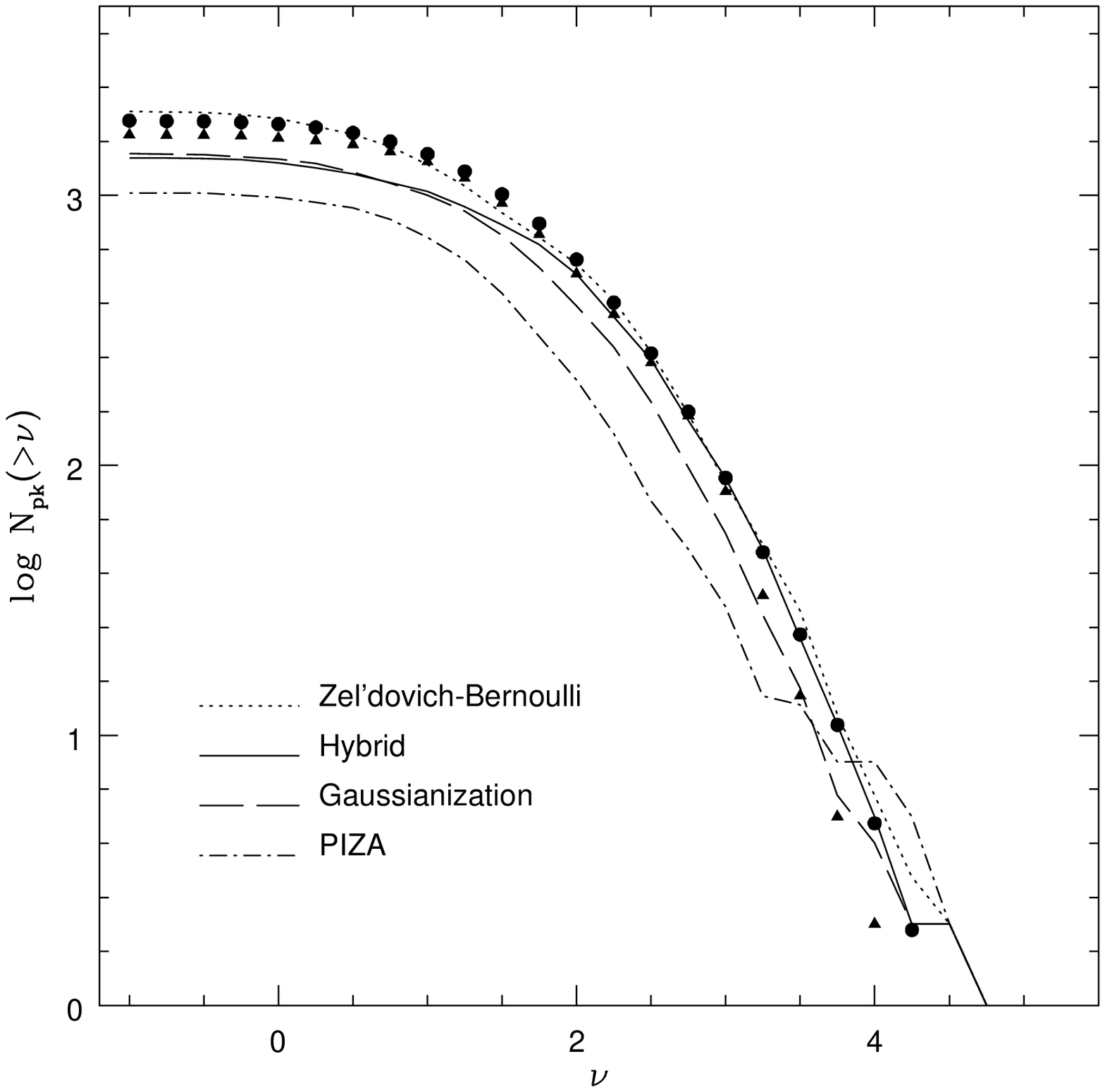}
}
\caption{ 
Same as Fig. 11, except that the initial density fields recovered by
 the \zb and PIZA schemes have been Gaussianized so that all the density
 fields shown in this Figure have a Gaussian PDF.	
}
\end{figure}

An obvious feature in the iso-density contour plots of Figure 1 
is that the structures in the reconstructed density fields appear more 
globular and isotropic compared to the corresponding structures in the 
true initial density field.
To investigate this quantitatively, we  define a  peak anisotropy parameter
 $\sigma_{a}$ to be
\be
\sigma_{a}^{2} = \sigma_{f}^{2} + \sigma_{e}^{2} + \sigma_{v}^{2}.
\label{eqn:sigadef}
\ee
In the above equation, $\sigma_{f},\  \sigma_{e},$ and  $\sigma_{v}$ are
the standard deviations in the density values of the pixels that share 
either a face, an edge, or a vertex with the peak pixel.
Figure 13 shows the \distrbn of the peak anisotropy parameters of all the 
peaks in the true and the reconstructed initial density fields.
We see that the median anisotropies of the peaks in all the reconstructed
fields are smaller than that of the true initial density field.
The anisotropy distributions of the hybrid and the \gau reconstructions 
are closest to that of the true initial distribution.
Linear theory and the Zel'dovich schemes recover a large tail of highly
anisotropic peaks, reinforcing our conclusions from Figure 1 
regarding the poor performance of these schemes near the high density regions.
The PIZA scheme recovers peaks that are more isotropic compared
 to the true peaks, and there are no peaks with $\sigma_{a} > 0.4$.
We also calculated this peak anisotropy \distrbn after imposing a Gaussian
PDF on all the reconstructed initial density fields.
We found that the distributions changed very little and our conclusions 
are unaffected by this.
We also found the same behavior for peaks above any given threshold value.
A plausible reason for this increased isotropy could be that, in the 
absence of any information about the initial small scale anisotropies in 
the final density field, which has been erased by non-linear evolution,
the \rec schemes tend to recover isotropic structures.

\begin{figure}
\centerline{
\epsfxsize=\hsize
\epsfbox[18 144 592 718]{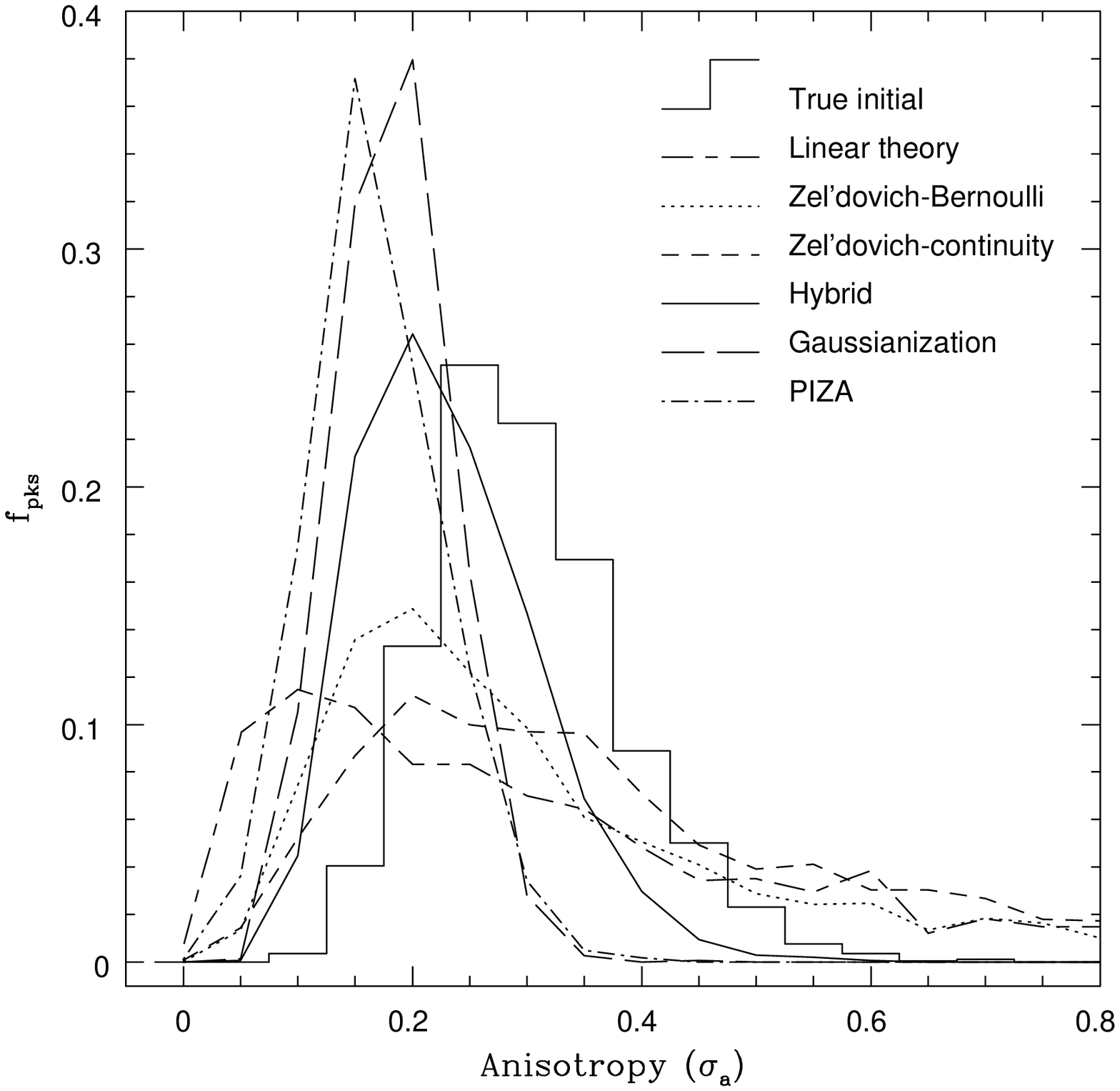}
}
\caption{ Distribution of the peak  anisotropies
(see equation~[\ref{eqn:sigadef}]) in the true and the reconstructed 
initial density fields.
}
\end{figure}

\section{DISCUSSION}
 
We have tested the accuracy of six different \rec schemes, which fall
into three general categories, for recovering  the smoothed
initial \den field from a gravitationally evolved mass 
distribution.
We compared their relative performances in an ideal setting in which 
the density fields are defined in a cubical box with
periodic boundary conditions.
We recovered the initial density fields smoothed with 
Gaussian filters of radii $R_{s} = 3, 5, 8 $ and $10h^{-1}$Mpc.
This range of smoothing lengths is representative of
 the smoothing scales at which the final density field can be reliably 
constructed from present day galaxy \z catalogs.
Our conclusions regarding the performance of the different \rec schemes
can be summarized as follows:
\begin{description}
\item[{(1)}] The linear theory recovered field is the worst match to 
the true initial field at all scales.
The fact that linear theory is so obviously bad compared to the other schemes
is quite encouraging, as it shows that the more sophisticated schemes
must be reasonably effective at undoing the effects of non-linear gravitational evolution. 
It also shows us that there is the potential to gain  better quality 
information by using reconstruction methods rather than applying simple 
linear theory analyses to observational data.

\item[{(2)}] The schemes based on the Zel'dovich approximation, namely
 the \zb and the \zc schemes, recover the \den field quite accurately in 
the quasi-linear regions, but they fail in the highly overdense regions 
where the mass \distrbn is very non-linear.
The recovery is generally quite poor for the smallest smoothing scale
considered (Gaussian smoothing with $R_{s} = 3h^{-1}$Mpc), when the 
perturbation theory assumptions underlying these schemes break down.
However, as expected, these reconstruction schemes become increasingly
accurate with larger smoothing, when the final density field is less
non-linear. 
Having said this, the Zel'dovich-Bernoulli scheme does perform unexpectedly 
well in some tests on small scales (see, e.g., Fig. 3 and Fig. 8). 
This is presumably due to the empirical correction described by 
equation~(\ref{eqn:vdndbb}).
The beneficial effects of this correction do not appear to be as great
 for large smoothings, when  the \zc scheme becomes relatively more accurate. 

\item[{(3)}] The \gau scheme recovers the initial density field quite 
robustly but not very accurately at all levels of smoothing.

\item[{(4)}] The hybrid method recovers the initial field both accurately 
and robustly.
There is no systematic failure even for mildly smoothed fields, and the 
distribution of peak statistics is recovered quite well.

\item[{(5)}] The PIZA scheme offers the best recovery of all the 
\rec schemes, at all smoothing scales.
It is able to reproduce the true initial density field on a
point-by-point basis very well even at the smallest smoothing scales
considered in this paper.
There is, however, too small a number of peaks in the reconstructed density 
field, and these recovered peaks also tend to be more isotropic than 
those in the true field.
\end{description}

All the \rec schemes in the first and the second category recover the 
initial density field from a  smoothed final density field that is
constructed using the galaxy positions in redshift catalogs.
A reconstruction from the final density {\it field} is quite convenient for
dealing with the effects of the selection function, and for applying local
 transformations to the  galaxy density field, which might, for example, 
include a model for the bias between galaxies and mass.
However, in the PIZA scheme, we recover the initial \den field using the 
{\it locations} of all the mass particles in the final mass \distrbn and then 
smooth this recovered initial field with a Gaussian filter 
before comparing it to the true smoothed initial density field.
We now describe a modification of the PIZA scheme that makes it 
applicable to smoothed final density fields.

Given a smoothed final density field on a grid, we
convert it to a particle distribution by placing in each grid 
cell a number of particles proportional to the density at that cell.
The locations of particles within the cell are chosen at random.
Since the PIZA scheme can be applied to large numbers of particles, we
choose a large enough number of particles that reduces the shot noise to
a negligible level (we represent a $3\hmpc$ smoothing volume at average 
density with $>400$ particles).
We then apply the PIZA procedure to this derived mass distribution,
 in the usual manner (described in \S2.6), and finally smooth this 
reconstructed initial density field.
We compare the smoothed initial density field reconstructed by this
 method to the true smoothed initial density field.
We do not describe the results in detail here but only state our main 
conclusions.
For a 3$h^{-1}$Mpc Gaussian smoothing of the final density field, 
the statistical tests of \S4 show the  \rec  to be only slightly 
inferior to that of the original PIZA scheme applied to the full mass 
distribution, and  still more accurate than the hybrid reconstruction 
method.
However, the recovered density field is very smooth, and there 
are fewer peaks compared even to the original PIZA reconstructed density 
field.
There are no peaks above $\nu > 2.5$, and the recovered peaks are
 all more isotropic than those in the true density field.

In carrying out the tests in this paper, we have assumed that we have complete
knowledge of the final real space mass distribution.
However, any attempt to recover the primordial density fluctuations in the 
local universe from galaxy \z catalogs, using 
the \rec methods we have discussed, will have to contend with three
important additional issues: the possibility of bias between  the galaxy 
distribution and the underlying mass distribution, the 
redshift space distortions arising from the peculiar velocities of galaxies,
 and the selection function of the survey itself.
A detailed examination and testing of solutions to these problems is beyond
the scope of this paper. 
However, we will briefly discuss some ways these effects can be handled.

\begin{description}
\item[{(1)}] Galaxy bias.
There is convincing evidence that the galaxy distributions selected
in the infrared and in the optical have different clustering 
properties (\cite{lahav90}; \cite{saunders92}; \cite{fisher94}).
Hence, it is only reasonable to assume that no one of the galaxy 
distributions is an unbiased tracer of the underlying mass distribution.
When the galaxy \distrbn is biased with respect to the mass distribution,
the observed galaxy number density fluctuations will not directly correspond
to the true mass density fluctuations.
However, it is reasonable to expect this relationship to be a local one, 
at least on or above one of the smoothing scales we have tested
(for an opposing point of view, see Bower et al.\ 1993).  
In this case, some of the different \rec schemes could deal
with bias between the galaxy and mass distributions.
For example, the \gau scheme can recover the initial mass density field 
from the
biased final \gal density field as long as biased galaxy formation  
preserves the rank order of the final mass density field.
This will be true if there is a local, monotonically increasing functional 
relationship between the mass and the galaxy density fields.
On the other hand, the two dynamical schemes that are based on the 
Zel'dovich approximation
require the final mass density field to compute the final gravitational
potential, although the \zb scheme can be used to reconstruct the 
primordial mass fluctuation field from the velocity 
potential, which can be derived directly from the peculiar velocity surveys
(see, e.g., Nusser, Dekel \& Yahil 1995).
The hybrid scheme can be applied to reconstruct biased \gal density 
fields, again assuming local biasing, using the procedure described in NW98.
In this, we first map the observed smoothed \gal \den field onto a final
PDF of the smoothed underlying mass distribution that is determined 
empirically assuming a value for the amplitude of mass density fluctuations.
This derived mass density field can then be evolved back in time to 
reconstruct the initial density field.
We can, in principle, use this mass density field in the PIZA scheme also, 
converting it to a particle-based realization as described above.

\item[{(2)}] Redshift distortions.
The second problem stems from the fact that the galaxy \distrbn in \z
space is a distorted version of the real space \gal distribution, because 
of the peculiar velocities of the galaxies (\cite{sargent77}; \cite{kaiser87}).
On small scales, the primary distortion is due to the velocity 
dispersions of clusters, which stretch compact real space clusters 
into elongated  ``Finger of God'' structures along the radial
direction in redshift space.
We can reduce this distortion by identifying all the galaxies
belonging to a cluster and collapsing them all to the \z of the 
cluster.
On large scales, coherent bulk inflows into overdense regions and 
outflows from underdense regions amplify the structures in the density
field in \z space. 
In the PIZA reconstruction scheme, it is a simple matter
to include these redshift distortions self-consistently (see CG97).
If the input particle positions are in redshift space, the peculiar
velocity component of the particle displacements can be automatically 
subtracted before computing the action. The PIZA method can therefore 
be used to recover an estimate of the real space final particle 
distribution from a redshift space final distribution, as well as
an estimate of the initial density field.
The other reconstruction schemes require a real space density field
as an input, which can be derived from the redshift space distribution
using the iterative procedure described 
in Yahil et al.\ (1991) and Gramann, Cen \& Gott (1994).
In this, we first derive a model for the peculiar velocity field from the 
\z space density field using the perturbation theory relations between
the two fields.
We then shift the location of a galaxy so that its redshift is now 
consistent with the Hubble flow and the peculiar velocity at its new 
location.
We derive a new density field from this corrected galaxy distribution
and use it to update our model of the peculiar velocity field.
We repeat these steps until the real space locations of the galaxies 
converge.
If the galaxy \distrbn is biased, we can first map the redshift space galaxy 
density field to a numerically determined real space mass density PDF before
deriving the peculiar velocity field (see \S4.1 in NW98 for a stepwise
description of this procedure).
We must bear in mind that it is necessary to assume a value for $\Omega$
in order to carry out any such procedures.

\item[{(3)}] The selection function.
Another problem in the analysis of observational data arises from the
decrease in the observed number density of galaxies with increasing 
distance from the observer in flux limited galaxy redshift surveys.
This requires that we smooth the final density field quite  heavily
to reduce the effects of shot noise, particularly if we wish to 
reconstruct over a large volume.
We also need to understand the selection function of the survey 
quite accurately, so that we can weight the galaxies in an optimal 
fashion, while deriving the final continuous density field.
There has been considerable progress of late in deriving
optimal estimators for the density field from the redshift survey 
data (\cite{springel98a} and references therein).
In dealing with the selection function, we must also be careful to explicitly
take into account the fact that it should be applied to the real space
positions of galaxies, in order to avoid the ``rocket effect'' 
(\cite{kaiser87}).

\end{description}

The primary goals for reconstructing the primordial density 
fluctuation field are twofold.
First, we can directly determine the statistical properties of the 
primordial density fluctuation field, such as, for example, the PDF
of the initial density fluctuations (\cite{ndy95}).
These properties can be used to constrain the theoretical 
models for the origin of these primordial fluctuations.
Secondly, we can evolve the reconstructed initial density field
forward in time using N-body methods and compare the evolved 
mass distribution with the observed galaxy  and cluster distributions, thereby 
leading to constraints on  the cosmological parameters and the galaxy 
formation models (W92; \cite{kolatt96}; NW98).

All the \rec methods that recover the primordial density fluctuations
from the density field 
in some volume require that we have an accurate map of the 
mass \distrbn in all the regions that have a significant gravitational 
effect on the mass \distrbn within this volume.
The \rec attempts so far have suffered from the lack of 
galaxy redshift catalogs that cover such  a large, representative 
region of the universe.
However, with the completion of new large \z surveys such as the PSCZ survey 
(\cite{pscz1}) and the ORS survey (ORS, Santiago et al. 1995, 1996), 
which map the \distrbn of galaxies over a large solid angle of the sky and 
to a reasonable depth, it is now becoming increasingly possible to 
accurately recover the properties of the initial density fluctuation field
 over a cosmologically interesting volume.
We hope that the comparative study of methods we have
undertaken in this paper will prove useful both to those undertaking
reconstruction analyses of these surveys and to those working on new, 
even more accurate ways of reversing the effects of gravity.

We thank David Weinberg for his suggestions and guidance.
This work was supported by NSF Grant AST-9616822 and
NASA Astrophysical Theory Grant NAG5-3111.

\end{document}